\documentclass[hidelinks]{eptcs}
\providecommand{\event}{} 

\usepackage{iftex}

\ifpdf
  \usepackage{underscore}         
  \usepackage[T1]{fontenc}        
\else
  \usepackage{breakurl}           
\fi

\usepackage{array}
\usepackage{amsmath}
\usepackage{amsfonts}
\usepackage{amsthm}
\usepackage{mathtools}
\usepackage{tikz}
\usepackage{sectsty}
\usepackage{MnSymbol,graphicx}
\usepackage[titletoc,toc,title]{appendix}
\usepackage{hyperref}
\usepackage{titlesec}
\usepackage{etoolbox}
\usepackage{todonotes}
\usepackage[hang,flushmargin]{footmisc}
\usepackage{xcolor}
\usepackage{comment}
\usetikzlibrary{automata, positioning, arrows}
\usepackage{setspace}

\newtheorem{theorem}{Theorem}[section]
\newtheorem{corollary}{Corollary}[section]
\newtheorem{lemma}{Lemma}[section]
\newtheorem{definition}{Definition}[section]
\newtheorem{example}{Example}[section]

\newcommand{\mydots}{{\scriptscriptstyle \ldots}}
\newcommand{\opt}[1]{#1?}
\newcounter{cont}

\definecolor{mDarkTeal}{HTML}{1C2833}

\newcommand{\ExternalLink}{%
    \tikz[x=1.2ex, y=1.2ex, baseline=-0.05ex]{%
        \begin{scope}[x=1ex, y=1ex]
            \clip (-0.1,-0.1) 
                --++ (-0, 1.2) 
                --++ (0.6, 0) 
                --++ (0, -0.6) 
                --++ (0.6, 0) 
                --++ (0, -1);
            \path[draw, 
                line width = 0.5, 
                rounded corners=0.5] 
                (0,0) rectangle (1,1);
        \end{scope}
        \path[draw, line width = 0.5] (0.5, 0.5) 
            -- (1, 1);
        \path[draw, line width = 0.5] (0.6, 1) 
            -- (1, 1) -- (1, 0.6);
        }
    }
\newcommand{\release}{v1.3}
\newcommand{\baseurl}{https://runtimeverification.github.io/vlsm-docs/\release/coqdoc/}
\newrobustcmd{\coqref}[2]{\href{\baseurl#1.html\##2}{\ExternalLink\hspace{-.1cm}}}

\newcommand{\nomessage}{     
\tikz[baseline]{
\draw[thin,shift={(0 cm,0.1cm)}] circle (1.5pt); 
\draw[thin,shift={(0 cm,0.1cm)}] (-0.09,0.09) -- (0.09,-0.09); 
\draw[thin,shift={(0 cm,0.1cm)}] (-0.09,-0.09) -- (0.09,0.09) }
} 

\makeatletter
\newcommand{\xdashrightarrow}[2][]{\ext@arrow 0359\rightarrowfill@@{#1}{#2}}
\newcommand{\xdashleftarrow}[2][]{\ext@arrow 3095\leftarrowfill@@{#1}{#2}}
\newcommand{\xdashleftrightarrow}[2][]{\ext@arrow 3359\leftrightarrowfill@@{#1}{#2}}
\def\rightarrowfill@@{\arrowfill@@\relax\relbar\rightarrow}
\def\leftarrowfill@@{\arrowfill@@\leftarrow\relbar\relax}
\def\leftrightarrowfill@@{\arrowfill@@\leftarrow\relbar\rightarrow}
\def\arrowfill@@#1#2#3#4{
  $\m@th\thickmuskip0mu\medmuskip\thickmuskip\thinmuskip\thickmuskip
   \relax#4#1
   \xleaders\hbox{$#4#2$}\hfill
   #3$
}
\makeatother

\newcommand{\transition}[5]{#2\xdashrightarrow[\ #3\ \rightarrow \ #5\ ]{#1} #4}
\newcommand{\transitionG}[5]{#2\xrightarrow[\ #3\ \rightarrow \ #5\ ]{#1} #4}
\newcommand{\transitionGT}[4]{\xrightarrow[\ #2\ \rightarrow \ #4\ ]{#1} #3}
\newcommand{\transitionV}[5]{#2\xRightarrow[\ #3\ \rightarrow \ #5\ ]{#1} #4}
\newcommand{\transitionVT}[4]{\xRightarrow[\ #2\ \rightarrow \ #4\ ]{#1} #3}

\newcommand{\valid}{\beta}
\newcommand{\Val}{\mathit{Validator}}
\newcommand{\localEquiv}{\mathit{localEqv}}
\newcommand{\globalEquiv}{\mathit{globalEqv}}
\newcommand{\localEquivocatorsFull}{\mathit{localEqv_{full}}}
\newcommand{\globalEquivocatorsFull}{\mathit{globalEqv_{full}}}

\newcommand{\messageDependencies}{message dependencies assumption}
\newcommand{\sender}{sender}
\newcommand{\senderM}[1]{\mathit{sender}(#1)}
\newcommand{\channelAuthentication}{channel authentication assumption}
\newcommand{\sent}{sent assumption}
\newcommand{\sentM}[2]{\mathit{sent}(#1,#2)}

\newcommand{\receivedM}[2]{\mathit{received}(#1,#2)}

\newcommand{\fullNode}{full node assumption}
\newcommand{\noMessageEquivocation}{\textit{no message equivocation constraint assumption\,}}
\newcommand{\MessageDependencies}{Message dependencies assumption}

\newcommand{\ChannelAuthentication}{Channel authentication assumption}
\newcommand{\Sent}{Sent assumption}
\newcommand{\Received}{Received assumption}
\newcommand{\HasBeenDirectlyObserved}{Directly observed information}
\newcommand{\HasBeenObserved}{Indirectly observed information}
\newcommand{\IndirectMessageDependencies}{Indirect message dependency relation}
\newcommand{\IncomparableMessages}{Incomparable messages}

\newcommand{\Eqv}{{\it eqv}}
\newcommand{\Proj}{\mathit{Proj}}
\newcommand{\len}{\mathit{len}}
\newcommand{\tr}{\mathit{tr}}

\newcommand{\N}{\mathbb{N}}
\newcommand{\concat}{\, +\hspace{-.1cm}+\, }
\newcommand{\send}{\mathit{send}}
\newcommand{\receive}{\mathit{receive}}
\newcommand{\nil}{[]}
\newcommand{\id}{{\it id}}
\newcommand{\obs}{{\it obs}}

\newcommand{\dependencies}{\mathit{dependencies}}
\newcommand{\receivedmessages}{\mathit{received\_messages}}
\newcommand{\sentmessages}{\mathit{sent\_messages}}
\newcommand{\messages}{\mathit{messages}}
\newcommand{\incomp}{\perp}

\newcommand{\weight}{\mathit{weight}}
\newcommand{\UMO}{\mathit{UMO}}
\newcommand{\MO}{\mathit{MO}}
\newcommand{\ELMO}{\mathit{ELMO}}
\newcommand{\umo}{\mathrm{UMO}}
\newcommand{\mo}{\mathrm{MO}}
\newcommand{\elmo}{\mathrm{ELMO}}
\newcommand{\state}[2]{\langle #1, #2\rangle}
\newcommand{\transitionU}[4]{#2\xrightarrow[#3]{#1} #4}

\newcommand{\const}{\mathit{c}}
\newcommand{\stateU}[2]{\langle #1, #2\rangle}

\theoremstyle{definition}
\newtheorem{assumption}{Assumption}
\newcommand{\emittable}{\mathit{emittable}}
\newcommand{\Emittable}{Emittable assumption}
\newcommand{\free}{\mathit{freeMsg}}
\newcommand{\psiVal}{\psi_{\mathit{msgValid}}}

\title{Validating Labelled State Transition and \\ Message Production Systems\\[.4em] {\Large A Theory for Modelling Faulty Distributed Systems}}

\author{
\centering
\begin{minipage}{13cm}
\begin{tabular}{ccc}
& Vlad Zamfir & \\
& {\footnotesize Ethereum Foundation} & \\
& {\footnotesize Runtime Verification, Inc.} & \\
\\
Mihai Calancea & Denisa Diaconescu & Wojciech Kołowski \\
{\footnotesize Runtime Verification, Inc.} & {\footnotesize University of Bucharest} & {\footnotesize Runtime Verification, Inc.} \\ 
 & {\footnotesize Runtime Verification, Inc.} &  \\ \\
Brandon Moore & Karl Palmskog & Traian Florin Șerbănuță \\
{\footnotesize Runtime Verification, Inc.} & {\footnotesize KTH Royal Institute of Technology} & {\footnotesize University of Bucharest} \\ 
 & {\footnotesize Runtime Verification, Inc.} & {\footnotesize Runtime Verification, Inc.}  \\ \\ 
Michael Stay & Dafina Trufas & Jan Tu\v{s}il \\
{\footnotesize Pyrofex Corp} & {\footnotesize University of Bucharest} & {\footnotesize Masaryk University} \\
 & {\footnotesize Runtime Verification, Inc.} & {\footnotesize Runtime Verification, Inc.} \\
\end{tabular}
\end{minipage}
}

\newcommand{\titlerunning}{Validating Labelled State Transition and Message Production Systems}
\newcommand{\authorrunning}{V. Zamfir et al.}
\begin{document}
\maketitle

\begin{abstract}
\textbf{Abstract.} Modeling and formally reasoning about distributed systems with faults is a challenging task. To address this problem, we propose the theory of Validating Labeled State transition and Message production systems (VLSMs). The theory of VLSMs provides a general approach to describing and verifying properties of distributed protocols whose executions are subject to faults, supporting a correct-by-construction system development methodology. 
The central focus of our investigation is equivocation, a mode of faulty behavior that we formally model, reason about, and then show how to detect from durable evidence that may be available locally to system components. Equivocating components exhibit behavior that is inconsistent with single-trace system executions, while also only interacting with other components by sending and receiving valid messages.  Components of system are called validators for that system if their validity constraints validate that the messages they receive are producible by the system. 
Our main result shows that for systems of validators, the effect that Byzantine components can have on honest validators is precisely identical to the effect that equivocating components can have on non-equivocating validators. Therefore, for distributed systems of potentially faulty validators, replacing Byzantine components with equivocating components has no material analytical consequences, and forms the basis of a sound alternative foundation to Byzantine fault tolerance analysis. All of the results and examples in the paper have been formalised and checked in the Coq proof assistant.
\end{abstract}

\newpage
\begin{spacing}{0.5}
\renewcommand{\contentsname}{Table of Contents\vspace{-0.6cm}}
\setcounter{tocdepth}{2}
\tableofcontents 
\end{spacing}

\section{Introduction}

The theory of Byzantine fault tolerance was conceived by Leslie Lamport \cite{lamport} for describing the problems that distributed systems face in the context of adversarial faults. Byzantine faulty components in a distributed system might behave in an arbitrary way in order to disrupt the operations of the system. Byzantine fault tolerance has long been considered to be the highest fault tolerance possible, because these faults can exhibit any behavior whatsoever. Satoshi Nakamoto relied on these concepts in order to describe the fault tolerance of Bitcoin, when using the concept of ``honest" and ``dishonest" nodes \cite{bitcoin}. However the notion that some nodes are honest and some nodes are Byzantine faulty is divorced from the economic realities of Bitcoin. In an economic model we would prefer to assume that every actor is able to modify the code of their component or somehow induce it to commit faulty behavior, but we would like to imagine that they would do it as a strategic choice. On the other hand, it may be very difficult to provide guarantees in a context where all components are strategically faulty and it may be realistic that only some nodes would be strategic, and therefore mixed models have been proposed \cite{bar}. 

In this paper, we propose the theory of \emph{Validating Labeled State transition and Message production systems} (VLSMs) as a fundamental approach to modeling and reasoning about distributed systems with faults. VLSMs were derived in the course of research on the CBC Casper consensus protocols \cite{vlad-2019}, as a tool for specifying both the protocols and the properties they should satisfy. 

The theory of VLSMs can be used for \emph{correct-by-construction} protocol design and implementation~\cite{ScienceProg}: we define an abstract class of protocols (satisfying some generic abstract properties), prove general results about protocols belonging to the class, and then obtain correct-by-construction protocols by instantiating the abstract components, or, alternatively, prove that concrete protocols satisfy those requirements.

VLSMs can formally describe and verify distributed protocols where faults occur during executions.  The central fault we investigate in this paper is that of {\em equivocation}. Equivocation refers to claiming different, irreconcilable beliefs, about a system to different parts of the system in order to steer protocol-abiding components into making inconsistent decisions \cite{clement,madsen,jaffe}. A key point is the fact that messages received from equivocating components seem to be valid messages. For example, if a system tries to come to a consensus about the value of a bit, an equivocating component may claim that the bit is $0$ to one part of the system, and $1$ to the other. Equivocation behavior cannot be produced by a single protocol execution, but only by multiple  protocol executions, i.e., an equivocating component behaves as if running multiple copies of the protocol.
We introduce a more general theory of equivocation, including {\em state equivocation} where a component splits off a parallel copy of itself, and {\em message equivocation} where components receive messages that have not been sent in the current trace of the system. 

In consensus protocols, it is common for components to ``validate'' the received messages in order to be sure that malformed messages are not received. We formalise this idea into a general formal notion of {\em validators} for a system. We are then able to show that the effect that Byzantine components can have on honest validators is no different than the effect equivocating components can have on non-equivocating validators.  This shows that equivocation fault tolerance analysis is a viable alternative to Byzantine fault tolerance analysis. This work opens the way for protocol designers to reason precisely about different types of faults and to budget for them separately, and thereby to create better consensus protocols than they could when they were budgeting only for Byzantine faults. 

All of the results and examples presented in the paper have been formalised and checked in the Coq proof assistant~\cite{VLSM13,Coq}.
We will use a symbol (\coqref{toc}{}) to refer to the corresponding formalisation in Coq. 

\section{VLSM Basic Notions}\label{vlsm}
In this section, we provide the basic definitions related to VLSMs and introduce some running examples.

\subsection{Main definition}

\begin{definition}[VLSM, \coqref{VLSM.Core.VLSM}{VLSM}]
\label{def:vlsm}
A \textbf{Validating Labeled State transition and Message production system} (\textbf{VLSM}, for short) is a structure of the form
$$\mathcal{V} = (L^{\mathcal{V}},S^{\mathcal{V}},S^{\mathcal{V}}_0,M^{\mathcal{V}}, M^{\mathcal{V}}_0, \tau^{\mathcal{V}}, \valid^{\mathcal{V}}),$$
where $L^{\mathcal{V}}$ is a set of \textit{labels}, $(S_0^{\mathcal{V}} \subseteq)\ S^{\mathcal{V}}$ is a non-empty set of \textit{(initial) states}, $(M_0^{\mathcal{V}} \subseteq)\ M^{\mathcal{V}}$ is a set of \textit{(initial) messages}, $\tau^{\mathcal{V}} : L^{\mathcal{V}} \times S^{\mathcal{V}} \times (\opt{M^{\mathcal{V}}}) \to S^{\mathcal{V}} \times (\opt{M^{\mathcal{V}}})$ is a transition function which takes as arguments a label, a state, and possibly a message, and outputs a state and possibly a message, while $\valid^{\mathcal{V}}$ is a \textit{validity constraint} predicate on the inputs of the transition function, i.e., $\valid^{\mathcal{V}} \subseteq L^{\mathcal{V}} \times S^{\mathcal{V}} \times (\opt{M^{\mathcal{V}}})$.\footnote{\label{note:option-message}For any set $M$ of messages, $\opt{M} = M \cup \{\nomessage\}$ is the extension of $M$ with $\nomessage$, where $\nomessage$ stands for {\em no-message}. We call a message {\em proper} if it is not $\nomessage$.} 
\end{definition}

When clear from the context, we will denote a VLSM $\mathcal{V} = (L^{\mathcal{V}},S^{\mathcal{V}},S_0^{\mathcal{V}},M^{\mathcal{V}}, M_0^{\mathcal{V}}, \tau^{\mathcal{V}}, \valid^{\mathcal{V}})$ simply by $\mathcal{V}$ and drop the superscripts on its elements. We will refer to a VLSM also as a {\em (VLSM) component}.  Similarly, we will denote an indexed family of VLSMs $\{\mathcal{V}_i = (L_i,S_i, S_{i,0}, M, M_{i,0}, \tau_i, \valid_i)\}_{i=1}^n$ simply by $\{\mathcal{V}_i \}_{i=1}^n$. 

In the sequel, let $\mathcal{V}$ be a VLSM. For a transition $\tau(l,s,m) = (s',m')$ in $\mathcal{V}$, we define the corresponding projections on states and messages by $\tau^s(l,s,m) = s'$ and $\tau^m(l,s,m) = m'$.

Let us emphasise our design choice to have a validity constraint predicate as part of a VLSM. As can be seen from the definition, the transition function is total; however, it indirectly becomes a partial function as the validity constraint predicate can filter out some of its inputs. We chose to specify transitions as a total function accompanied by the validity constraint predicate as we aim to allow protocol designers to envision all possible dynamics of their protocol and to have the power to point out which are the ideal/protocol abiding dynamics. The set of labels in a VLSM implies a deterministic behavior; however, it is possible to have multiple parallel transitions between two states, each with their own label. Consequently, in a VLSM we can have three types of transitions, states, messages, and traces.

For any label $l$, state $s$, and message $m$ in $\mathcal{V}$, we always have the transition $\tau(l,s,m) = (s',m')$ using $l$, $s$, and $m$,  and we denote it  by $\transition{l}{s}{m}{s'}{m'}$. A \textit{trace} is a sequence of transitions starting in an initial state.

\subsection{Constrained states and messages}

If the validity constraint predicate $\valid(l,s,m)$ holds for a label $l$, a state $s$, and a message $m$ in $\mathcal{V}$, the transition $\tau(l,s,m) = (s',m')$ is called \textit{constrained} and we denote it by $\transitionG{l}{s}{m}{s'}{m'}$.  When clear from the context, we will refer to constrained transitions simply as transitions. A \textit{constrained trace} (\coqref{VLSM.Core.PreloadedVLSM}{finite_constrained_trace_init_to_direct}) is a trace in which all transitions are constrained. A \textit{constrained state} (\coqref{VLSM.Core.PreloadedVLSM}{constrained_state_prop_direct}) is a state that appears in a constrained trace, while a \textit{constrained message} (\coqref{VLSM.Core.PreloadedVLSM}{constrained_message_prop_direct})  is a message emitted on a constrained trace. More formally,

\begin{definition}[Constrained states and messages]
The \textbf{constrained states} and \textbf{constrained messages} associated with $\mathcal{V}$ are  $S^{\const} = \bigcup_{n=0}^\infty S_{n}^{\const}$  and  $M^{\const} = \bigcup_{n=0}^\infty M_{n}^{\const}$, where 
\begin{align*}
	S^{\const}_{0} &= S_0 \mbox{ and }M^{\const}_{0} = M_0\cup \{\nomessage\}, \\
	S^{\const}_{n+1} &= S^{\const}_{n}\ \cup \{\tau^s(l,s,m)\, \ |\  l \in L,  \ s \in S^{\const}_{n},\ m \in M,\  \valid(l,s,m)\}, \\
	M^{\const}_{n+1} &= M^{\const}_{n} \cup \{\tau^m(l,s,m)\ |\  l \in L,\ s \in S^{\const}_{n},\ m \in M,\  \valid(l,s,m)\}.
\end{align*}
\end{definition}

\subsection{Valid states and messages}

A \textit{valid trace} (\coqref{VLSM.Core.PreloadedVLSM}{finite_valid_trace_init_to_from_constrained}) is a constrained trace in which any input message is a \textit{valid message} (\coqref{VLSM.Core.PreloadedVLSM}{option_valid_message_prop_from_constrained}), i.e., it is either $\nomessage$, an initial message, or one emitted on some (potentially different) valid trace. \textit{Valid transitions} are transitions appearing in some valid trace. We denote a valid transition $\tau(l,s,m) = (s',m')$ by
$\transitionV{l}{s}{m}{s'}{m'}.$
Intuitively, valid traces are executions of a VLSM in isolation, in which the only acceptable inputs are the ones the VLSM can produce by itself (in some other valid execution).
In contrast, constrained traces are executions of a VLSM in an arbitrary environment, in which any message is acceptable as an input.
A \textit{valid state} (\coqref{VLSM.Core.PreloadedVLSM}{valid_state_prop_from_constrained}) is a state that appears in a valid trace. Formally,

\begin{definition}[Valid states and valid messages, \coqref{VLSM.Core.VLSM}{valid_state_message_prop}]\label{states-messages}
The \textbf{valid states} and \textbf{messages} associated with $\mathcal{V}$ are  $S^{v} = \bigcup_{n=0}^\infty S_n^{v}$ and $M^{v} = \bigcup_{n=0}^\infty M_n^{v}$, where
\vspace{-.2cm}
\begin{align*}
	S^{v}_{0} &= S_0  \mbox{ and } M^{v}_{0} = M_0 \cup \{\nomessage\}, \\
	S^{v}_{n+1} &= S^{v}_{n}\ \cup \{\tau^s(l,s,m)\, \ |\  l \in L,\ s \in S^{v}_{n},\ m \in M^{v}_{n},\ \valid(l,s,m) \}, \\
	M^{v}_{n+1} &= M^{v}_{n} \cup \{\tau^m(l,s,m)\ |\  l \in L,\ s \in S^{v}_{n},\ m \in M^{v}_{n},\ \valid(l,s,m) \}.
\end{align*}
\end{definition}

In general, the question whether a state or a message is valid is undecidable. For example, take the set of states to be the configurations of a Turing machine and consider as constrained transitions the updates of the Turing machine's state. Since the halting problem is undecidable, we cannot know for an arbitrary Turing machine whether it will reach that state.  We can also say that the VLSM emits a valid message if and only if the Turing machine halts, which makes the question of whether a message is a valid one undecidable in general.  However, we often make extra assumptions about the shape of states and messages that enable us to decide this problem.

\subsection{Example: VLSMs performing multiplication of inputs}\label{subsec:multiplication}

Let us consider the following VLSM  ${\mathcal{P}_2}$ which \textit{receives any integer and emits the double of it} (\coqref{VLSM.Examples.Tutorial.Multiply}{doubling_vlsm}). Formally, ${\mathcal{P}_2}$ has only one label (which can be omitted when clear from the context) $L^{\mathcal{P}_2} = \{d\}$,  the set of states is the set of integers $S^{\mathcal{P}_2} = \{n \ |\ n \in \mathbb{Z}\}$, the initial states are the natural numbers greater or equal to $2$, $S_0^{\mathcal{P}_2} = \{ n\ |\ n \geq 2 \}$, the messages are the integers $M^{\mathcal{P}_2} = \mathbb{Z}$, and only one initial message $M_0^{\mathcal{P}_2} = \{2\}$. For any integers $n$ and $i$, we have the following transition in ${\mathcal{P}_2}$ 
		$${\transition{d}{n }{\ \, i}{n-i}{2i\ }},$$
while the validity constraint predicate in ${\mathcal{P}_2}$  is defined as
		$${\valid^{\mathcal{P}_2} = \{(d,n, i)\ |\ n \geq i \geq 2\}}.$$
Let us explore how all the above terminology translates into this example.
\begin{description}

\item[Transitions.] For example, let us consider $\transition{}{4 }{\ 10}{-6}{20_{}}$. This clearly is a transition in the VLSM ${\mathcal{P}_2}$, however it is not a constrained transition as $4 \not\geq 10$.

\item[Constrained transitions.]  For example,  we can see that $\transitionG{}{5}{3}{2}{6}$  is a constrained transition as the input for this transition satisfies the validity constraint predicate. However, it is not a valid transition as the message  $3$ cannot be produced by any constrained transition.

\item[Constrained traces.] Let us consider the sequence
		$\transitionG{}{5}{3}{2}{6} 
		     \transitionGT{}{2}{0}{4}  
		  $. This is a sequence of constrained transitions starting from the initial state $5$.

\item[Constrained states.] The constrained states associated with ${\mathcal{P}_2}$ are all states of the form $\{ n \ |\ n \geq 0\}$. This can be easily seen since constrained states can be reached from initial states  by a sequence of constrained transition in which we can only subtract numbers greater or equal to $2$ from the state such that we cannot obtain a negative number. Note that $0$ and $1$ can be obtained via this process.

\item[Constrained messages.] The constrained messages associated with ${\mathcal{P}_2}$ are multiples  of $2$ (excluding $0$).  The transition function produces even numbers as output messages. Since constrained transitions use only positive numbers greater that $2$ as inputs, they can only produce natural numbers as output messages and $0$ cannot be produced.		  

\item[Valid transitions.] For example, $\transitionV{}{8}{2}{6}{4} $ is a valid transition. Clearly, the validity constraint predicate is satisfied for the input $(d,8,2)$, $8$ is a valid state as all initial states are valid, and the message $2$ is valid as all initial messages are valid.
		  
\item [Valid trace.] For example, the sequence
		$\transitionV{}{8}{4}{4}{8} 
		     \transitionVT{}{2}{2}{4}  
		     \transitionVT{}{2}{0}{4_{}}  
		  $
 is a valid trace. All these transitions are constrained and, moreover, they involve only valid states and messages.

\item[Valid states.] The valid states associated with ${\mathcal{P}_2}$ coincide with the constrained states, i.e., states of the form $\{ n \ |\ n \geq 0\}$. 

\item[Valid messages.] The valid messages associated with ${\mathcal{P}_2}$ are powers of $2$ (excluding $1$). Valid messages are a subset of constrained messages, so valid messages must be even natural numbers. Since the only initial message is $2$ and valid messages are those produced by constrained transitions using only valid messages, we get that valid messages are the powers of $2$. Note that $1$ cannot be emitted by any transition.
\end{description}

The above VLSM can be easily generalized for any prime number $p \geq 2$ (\coqref{VLSM.Examples.Tutorial.PrimesComposition}{radix_vlsm}). Namely, we can define the VLSM ${\mathcal{P}_p}$ with the same ingredients as above, except for
\begin{center}
$M_0^{\mathcal{P}_p} = \{p\}$  \quad and \quad $\transition{}{n }{\ \, i}{n-i}{p\cdot i}$.
\end{center}
The constrained and valid states associated with ${\mathcal{P}_p}$ coincide again, i.e., $\{ n \ |\ n \geq 0\}$, the constrained messages are multiples of $p$ (excluding $0$), while valid messages are powers of $p$ (excluding $1$).

\subsection{Example: Unvalidating Message Observer (UMO) components}\label{UMO}

UMO is an example of a VLSM which records all the transitions that it performs. An UMO component has two labels, $\send$ and $\receive$. States are tuples of the form $ \stateU{o}{i}$, where $i$ is an address\footnote{For simplicity, we choose addresses within the set of natural numbers $\N$.} and $o$ is a list of message observations.  A message observation is of the form $(\send, m)$ or $(\receive, m)$, where $m$ is a message. Messages are  states. When an UMO component sends a message, it sends its current state; therefore, sometimes it is easier to think of states as messages. Formally, we define the set of states $S = \bigcup_{n=0}^\infty S_{n}$, where \footnote{$\nil$ stands for the empty list, while $\concat$ stands for the usual concatenation of lists.}  
\begin{align*}
S_{0} &= \{\stateU{\nil}{i}\mid i\in \N\}, \\
S_{n+1} &= S_{n} \cup \bigcup_{\substack{\stateU{o}{i}\, \in\, S_{n} \\ \stateU{o'}{i'}\, \in\, S_{n} }} \{\stateU{o \concat [(\send, \stateU{o'}{i'})]}{i}, \stateU{o \concat [(\receive, \stateU{o'}{i'})]}{i}\}.
\end{align*}

For any state $\stateU{i}{o}$, we define the following operations: 
\begin{align*}
\obs(\stateU{o}{i}) &= o, \\ 
\id(\stateU{o}{i}) &= i, \\
\sentmessages(\stateU{o}{i}) &= \{m\ |\ (\send,m) \in o\}, \\ 
\receivedmessages(\stateU{o}{i}) &= \{m\ |\ (\receive,m) \in o\}, \\ 
\messages(\stateU{o}{i}) &= \sentmessages(\stateU{o}{i}) \cup \receivedmessages(\stateU{o}{i}).
\end{align*}

The \textit{UMO component of address $i \in \N$} (\coqref{VLSM.Examples.ELMO.UMO}{UMO_component}) is the VLSM $\mathcal{U}_i$ with $L = \{\send,\receive\}$, $S$ defined as above, $S_0^i = \{\stateU{\nil}{i}\}$, $M = S$, and $M_0 = \emptyset$, The transition function of $\mathcal{U}_i$ defined as
\begin{align*}
\tau(\send,s,\nomessage) &= (\stateU{\obs(s) \concat [(\send, s)]}{\id(s)}, s), \\
\tau(\receive,s, m) &= (\stateU{\obs(s) \concat [(\receive,m)]}{\id(s)}, \nomessage), \\
\tau(l,s,m) &= (s,\nomessage), \mbox{ in any other case},
\end{align*}
while the validity constraint predicate is defined as
$$\valid_{\UMO}(l,s,m) = ((l = \send) \wedge (m = \nomessage)) \vee ((l = \receive) \wedge (m \neq \nomessage)).$$

For simplicity, we denote a constrained transition in $\mathcal{U}_i$ by {$\transitionU{l}{s}{m}{s'}$}, where $l$ can be either $\send$ (in which case the transition is $\tau(\send,s,\nomessage) = (s',m)$ and $m = s$) or $l$ can be $\receive$ (in which case the transition is $\tau(\receive,s,m) = (s',\nomessage)$). For example, the following is a constrained trace in the UMO component of address $2$; note that the last state  is not a valid state as message $m_2$ cannot be emitted on any valid trace.
$$
 \stateU{[]}{2}
\xrightarrow[m_1 = \stateU{[]}{2}]{\send}
\stateU{[{(\send, m_1)}]}{2}  
\xrightarrow[m_2 = \stateU{[(\send, \stateU{[]}{1}), (\send, \stateU{[]}{2})]}{1} ]{\receive} 
\stateU{[(\send, m_1), (\receive, m_2)]}{2}
$$

By definition, for any constrained state there is a constrained trace leading to it. However, what is additional for UMO components is that in any constrained state of an UMO component, there is a unique trace encoded into that state which leads to it.

\begin{lemma}[\coqref{VLSM.Examples.ELMO.UMO}{constrained_state_contains_unique_constrained_trace}]\label{UMO-component-trace}
From every constrained state of an UMO component we can extract a unique constrained trace reaching it.
\end{lemma}

\section{Composition of VLSMs} \label{composition}

A single VLSM represents the local point of view of a component in a distributed system. We can obtain a global point of view of a system by combining multiple VLSMs and lifting their local validity constraints.  In the sequel, whenever we consider a family of VLSMs we assume that they have the same set of messages.
Let  $\{\mathcal{V}_i \}_{i=1}^n$ be an indexed set of VLSMs over the same set of messages $M$.

\subsection{Free composition}
The most natural way of putting together VLSM components is via a {\em free composition} in which we consider the product of their states and let the global transition function and the global validity constraint predicate be defined component-wise, guided by labels belonging to individual components. Formally, we have the following definition.

\begin{definition}[Free composition, \coqref{VLSM.Core.Composition}{free_composite_vlsm}]
The \textbf{free VLSM composition} of $\{\mathcal{V}_i \}_{i=1}^n$ is the VLSM
$${ \sum_{i=1}^n}\, \mathcal{V}_i = (L,S, S_0, M, M_0, \tau, \valid),$$ where
$L = {\bigcup_{i=1}^n \{i\} \times L_i}$ is the disjoint union of labels,
$S = {\prod_{i=1}^n S_i}$ is the product of states,
$S_0 = {\prod_{i=1}^n S_{i,0}}$ is the product of initial states,
$M$ is the same set of messages as for each $\mathcal{V}_i$,
$M_0 = { \bigcup_{i=1}^n M_{i,0}}$ is the union of all initial messages,
$\tau: L \times S \times M? \rightarrow  S \times M?$  and $\valid \subseteq L \times S \times M?$ are defined component-wise, guided by labels, i.e.,
\begin{center}
$\tau(\langle j, l_j \rangle, \langle s_1,  \mydots s_n\rangle, m) = 
		\left(\langle s_1,\mydots, s_{j-1}, \tau^s_j(l_j, s_j, m), s_{j+1}, \mydots s_n\rangle, \tau^m_j(l_j, s_j, m)\right),$
		
$\valid(\langle j,l_j \rangle,\langle s_1,\mydots,s_n\rangle,m) = \valid_j(l_j,s_j,m).$
\end{center}
\end{definition}

We will refer to the states of a free composition as {\em composite states}. Constrained traces in a component can be naturally "lifted" to the free composition.

\begin{lemma}[\coqref{VLSM.Core.Composition}{lift_to_composite_VLSM_embedding}]\label{lem:lift-trace}
   Let $s_0\xrightarrow[m_1\ \rightarrow \ m'_1]{l_1} s_1 \xrightarrow[m_2\ \rightarrow \ m'_2]{l_2} \cdots \xrightarrow[m_n\ \rightarrow \ m'_n]{l_n} s_n$ be a constrained trace in $\mathcal{V}_i$ and let $\sigma_0$ be an initial state of the free composition $\mathcal{V}$ such that 
   the $i^{th}$ component of $\sigma_0$ is $s_0$.
    Then $$\sigma_0\xrightarrow[m_1\ \rightarrow \ m'_1]{\langle i, l_1 \rangle} \sigma_1 \xrightarrow[m_2\ \rightarrow \ m'_2]{\langle i, l_2\rangle} \cdots \xrightarrow[m_n\ \rightarrow \ m'_n]{\langle i, l_n \rangle} \sigma_n$$  is a constrained trace in $\mathcal{V}$, where for all $1 \leq j \leq n$, $\sigma_j$ is equal on all components to $\sigma_0$, except for its $i^{th}$ component, which is $s_j$. 
\end{lemma}

The free composition allows messages produced by one VLSM to be received by any other VLSM, including itself.  However,  note that a VLSM may receive a message that was not sent earlier in a trace.  We address the issue of receiving messages produced on alternative traces in the sections dedicated to equivocation.

\subsection{Constrained composition}\label{composition:constrained}

The validity constraint in the free composition lift globally the local validity constraint of each component involved in  the composition. Protocol designers may want to impose further global restrictions, stronger than the ones that can be specified locally on individual components; we capture this phenomenon by the notion of a {\em composition constraint} which can be enforced further in a free composition.

\begin{definition}[Constrained composition, \coqref{VLSM.Core.Composition}{composite_vlsm}]\label{constraint-composition}
A \textbf{composition constraint} $\varphi$ is a predicate additionally filtering the inputs for the composed transition function, 
$\varphi \subseteq L \times S \times M?$.
The \textbf{constrained VLSM composition under $\varphi$} of $\{\mathcal{V}_i\}_{i=1}^n$ is the VLSM which has the same components as the free composition, except for the validity predicate which is further constrained by $\varphi$, namely\footnote{For a more intuitive notation, for any $\beta, \varphi \subseteq L \times S \times M?$, we denote by $\beta \wedge \varphi$ the set intersection $\beta \cap \varphi$.}
$$\Bigr({\sum_{i=1}^n} \mathcal{V}_i \Bigr) \Bigr|_\varphi = (L,S, S_0, M, M_0,\tau, \valid \wedge \varphi).$$
\end{definition}

Note that when a composition constraint $\varphi$ coincides with $L \times S \times M?$, then the two notions of composition coincide.
We  emphasize that a constrained composition of a family of VLSMs can have fewer valid composite states/messages than the free composition of that family.

Let $\mathcal{V} =\ \bigr({\sum_{i=1}^n} \mathcal{V}_i\bigr)\bigr|_\varphi$ be a constrained composition. Given a transition in $\mathcal{V}$ of the form
$$\transition{\langle j,l_j\rangle}{\langle s_1,\mydots,s_j,\mydots,s_n\rangle}{m}{\langle s_1,\mydots,s_j',\mydots,s_n\rangle}{m'},$$
its {\em (transition) projection} (\coqref{VLSM.Core.ProjectionTraces}{composite_transition_item_projection}) on component $j$ is
$$\transition{l_j}{s_j}{m}{s_j'}{m'}.$$
Given a trace $\tr$ in $\mathcal{V}$, its {\em (trace) projection} (\coqref{VLSM.Core.ProjectionTraces}{finite_trace_projection_list}) on component $j$ consists of all the projections of the transitions from $\tr$ with labels of the form $\langle j,l_j\rangle$ taken in the same order as  in $\tr$.

Furthermore, we can extract traces from those in $\mathcal{V}$ which remove the actions of components whose index is not in a fixed set of indices $I$. Indeed, let $\tr$ be a valid trace in $\mathcal{V}$. We construct a trace $\tr_{|_I}$ from $\tr$ by eliminating all transitions with labels of the form $\langle j,l_j\rangle$ with $j \not \in I$ and, for any composite state appearing in $\tr$, we eliminate all states not corresponding to the indices $I$, i.e., we consider the corresponding state in  $\mathcal{V}_I = { \sum_{\substack{i=1\\i\in I}}^n}\, \mathcal{V}_i$. We denote the set of {\em projected traces for indices $I$} by $\mathit{Tr}(\mathcal{V})_{|_I} = \{\tr_{|_I}\ |\ \tr \mbox{ is a valid trace in } \mathcal{V}\}$.

\begin{definition}[Induced projection, \coqref{VLSM.Core.ProjectionTraces}{composite_vlsm_induced_projection}]\label{induced-projection}
For any $j \in \{1,\mydots, n\}$, the \textbf{induced  $j^{th}$ projection} of $\mathcal{V}$ is the VLSM
$$\Proj_j(\mathcal{V}) = (L_{j}, S_{j}, S_{j,0}, M, M^{v},\tau_{j}, \valid_{j}),$$
where $L_{j}, S_{j}, S_{j,0}, M, \tau_{j}, \valid_{j}$ are the same as in the original component $\mathcal{V}_j$, and the set of initial messages is the set of all valid messages of the composition $\mathcal{V}$.
\end{definition}

The induced $j^{th}$ projection and $\mathcal{V}_j$ do not usually coincide. While $\mathcal{V}_j$ accepts as valid messages only those that can be produced by itself, $\Proj_j(\mathcal{V})$ accepts as valid messages all valid messages of the composition.
The projections of the valid traces from $\mathcal{V}$ are valid in their corresponding induced projections.
Due to the constrained composition, there can be valid traces in the induced projections which cannot be lifted to valid traces in the constrained composition. 
However, valid traces in the composition (under the same constrained composition) of the original components and of the induced projections coincide.

\subsection{Example: Composition of VLSMs performing multiplication of inputs}

Let us consider the free composition $\mathcal{P}_2 + \mathcal{P}_3$ (\coqref{VLSM.Examples.Tutorial.Multiply}{free_composition_23}) of $\mathcal{P}_2$ and $\mathcal{P}_3$ introduced in Subsection \ref{subsec:multiplication}. A composite state in $\mathcal{P}_2 + \mathcal{P}_3$ is of the form $\langle n_1,n_2 \rangle$ with $n_1,n_2\in \mathbb{Z}$. 
The valid messages in $\mathcal{P}_2 + \mathcal{P}_3$ are 
$$M^{\mathcal{P}_2 + \mathcal{P}_3}_{v} = \{2^i \cdot 3^j\ |\ i+j \geq 1\}.$$ 
Note that $2,3 \in M^{\mathcal{P}_2 + \mathcal{P}_3}_{v}$, while $1 \not\in M^{\mathcal{P}_2 + \mathcal{P}_3}_{v}$.  

Now let us consider the following composition constraint which guarantees that the received messages are even numbers (\coqref{VLSM.Examples.Tutorial.Multiply}{parity_composition_23})
\begin{align*}
\varphi_{\mathit{parity}} & = \{(\langle \mathcal{P}_2,d\rangle,\langle n_1,n_2\rangle, i)\ |\  i\ \mathrm{mod}\ 2 = 0 \}\  \cup 
 \{(\langle \mathcal{P}_3,d\rangle,\langle n_1,n_2\rangle, i)\ |\  i\ \mathrm{mod}\ 2 = 0 \}.
\end{align*}
The valid messages in the composition of $\mathcal{P}_2$ and $\mathcal{P}_3$ constrained by $\varphi_{\mathit{parity}}$ are (\coqref{VLSM.Examples.Tutorial.Multiply}{parity_composition_23_valid_messages_powers_of_23})
$$M^{(\mathcal{P}_2 + \mathcal{P}_3)|_{\varphi_{\mathit{parity}}}}_{v} = \{2^i \cdot 3^j\ |\ i \geq 1\} \cup \{3\}.$$
Note that $3$ cannot be received or emitted because of the composition constraint, but it is a valid message as it is an initial message in $\mathcal{P}_3$.

\subsection{Example: The UMO protocol}

Let $\{\mathcal{U}_i\}_{i=1}^n$ be a set of UMO components (Subsection \ref{UMO}) indexed by their addresses, i.e, $\mathcal{U}_i$ is the UMO component of address $i$. The \textbf{UMO protocol} $\umo(\mathcal{U}_i)_{i=1}^n$ is defined as the free VLSM composition of $\{\mathcal{U}_i\}_{i=1}^n$ (\coqref{VLSM.Examples.ELMO.UMO}{UMO}).

The following result allows us to recover a trace from a constrained composite state of an UMO protocol by combining the traces extracted by Lemma \ref{UMO-component-trace} from each component of the composite state. However, from any constrained composite state we can extract more traces leading to it, depending how we combine the traces leading to the components of the composite state. 

\begin{lemma}[\coqref{VLSM.Examples.ELMO.UMO}{finite_valid_trace_from_to_UMO_state2trace}]\label{UMO-protocol-trace}
From every constrained state of an UMO protocol we can extract a constrained trace reaching it.
\end{lemma}

As a particular case, let us consider the UMO components $\mathcal{U}_1$, $\mathcal{U}_2$, and $\mathcal{U}_3$ of address $1$, $2$, and $3$, respectively. The following is a constrained trace of $\umo(\mathcal{U}_i)_{i=1}^3$:
\begin{align*}
\begin{bmatrix}
\state{[]}{1} \\
\state{[]}{2} \\
\state{[]}{3} \\
\end{bmatrix}
\xrightarrow[m_1 = \state{[]}{1}]{\langle 1, \send \rangle}
\begin{bmatrix}
\state{[(\send,m_1)]}{1} \\
\state{[]}{2} \\
\state{[]}{3} \\
\end{bmatrix}
\xrightarrow[m_2 = \state{[]}{2}]{\langle 2, \send \rangle}
\begin{bmatrix}
\state{[(\send,m_1)]}{1} \\
\state{[(\send,m_2)]}{2} \\
\state{[]}{3} \\
\end{bmatrix}
\xrightarrow[m_1]{\langle 3, \receive \rangle}
\begin{bmatrix}
\state{[(\send,m_1)]}{1} \\
\state{[(\send,m_2)]}{2} \\
\state{[(\receive,m_1)]}{3} \\
\end{bmatrix}
\end{align*}

Now suppose we want to define a refinement of the above UMO protocol in which $\mathcal{U}_1$ can receive messages only from $\mathcal{U}_2$, $\mathcal{U}_2$ can receive messages only from $\mathcal{U}_3$, and $\mathcal{U}_3$ can receive messages only from $\mathcal{U}_1$. We can obtain this (global) restriction using the following composition constraint:
\begin{align*}
\varphi = \{(\langle i,\receive \rangle,\sigma, m)\ |\ \sigma \in S \times S \times S,\ m \in S,\ \id(m) = (i\ \mathrm{mod}\ 3) + 1\} \cup \{(\langle i,\send \rangle,\sigma, m)\}. 
\end{align*}

\section{Validators}

In a distributed system we are interested in dealing with valid messages. Sometimes a single component does not have the capabilities to filter out  junk or malformed information. Validators are components strong enough to locally guarantee that a message is valid. A validator enforces a (global) composition constraint locally: if a transition would cause its view of the system to violate the global constraint, then it will not be a constrained transition in the component.
For example, the UMO components introduced in Subsection \ref{UMO} cannot establish if a message is valid. Indeed, let us consider the following constrained transition in the UMO component $\mathcal{U}_2$:
\[
s
\xrightarrow[m = \stateU{[(\send, \stateU{[]}{1}), (\send, \stateU{[]}{2})]}{1} ]{\receive} 
s'
\]
Message $m$ cannot be emitted from any composite state of the UMO protocol $\umo(\mathcal{U}_i)_{i=1}^3$, as it is ruled out by the definition of the transition function. Hence this is not a valid message in the free composition.

\subsection{Definition of validator}

Let $\mathcal{V} =\ \bigr({\sum_{i=1}^n} \mathcal{V}_i\bigr)\bigr|_\varphi $ be the composition under $\varphi$ of $\{\mathcal{V}_i\}_{i=1}^n$. 
Let $j \in \{1,\mydots,n\}$ be the index of a component.

\begin{definition}[Validator, \coqref{VLSM.Core.ProjectionTraces}{component_projection_validator_prop}]
The component $\mathcal{V}_j$ is a \textbf{validator} for $\mathcal{V}$ if any constrained transition from a constrained state in $\mathcal{V}_j,$
$s_j\xrightarrow[m\ \rightarrow \ m']{l} s_j',$
can be "lifted" to a valid transition in $\mathcal{V},$
$\sigma\xRightarrow[m\ \rightarrow \ m']{\langle j,l\rangle} \sigma',$
such that the $j^{th}$ components of $\sigma$ and $\sigma'$ are $s_j$ and $s_j'$, respectively.
\end{definition}

We would like to establish generic conditions for a component to be validating for
a free composition it is part of.
We make the following assumptions:

\begin{assumption}[Channel Authentication, \coqref{VLSM.Core.Equivocation}{channel_authentication_prop}]
\label{channel-authentication}
Every message must have an identifiable sender. We assume there exists a mapping $\sender$ from each message to its corresponding \sender. 
\end{assumption}

\begin{assumption}[Message Dependencies, \coqref{VLSM.Core.MessageDependencies}{MessageDependencies}]
\label{message-dependencies}
 There exists an oracle $\dependencies$ which, given a message, determines its direct dependencies. 
We assume that if a message is a constrained message of the component (i.e., the component can emit it), then the component can emit it while only relying on its direct dependencies.
This assumption also induces a dependency relation between messages (\coqref{VLSM.Core.MessageDependencies}{msg_dep_rel}). 
We further assume that $\dependencies$ yields a finite set for any message, hence the dependency relation induced by it is well-founded (\coqref{VLSM.Core.MessageDependencies}{msg_dep_happens_before_wf}), and that messages emitted by the same component on the same run of the protocol are ordered by this relation (\coqref{VLSM.Core.MessageDependencies}{has_been_sent_msg_dep_comparable_prop}).
\end{assumption}

\begin{assumption}[Emittable, \coqref{VLSM.Core.Validators.FreeCompositionValidator}{emittable}]
\label{emittable}
There exists an oracle $\emittable$ which can tell whether a message is a constrained message for any of the components $\mathcal{V}_i$.
\end{assumption}

Under the above assumptions, let $\psi_\free(m)$ be a predicate which characterizes whether the message $m$ is 
valid for the free composition:
\begin{align*}
\psi_\free(m) &\ =\ \sender(m) \in \{1, \mydots, n\}\ \wedge\ \emittable_{\sender(m)}(m)\ \wedge \\
 & \hspace{1cm}  \forall m' (m' \in \dependencies(m) \to \psi_\free(m')).
\end{align*}

\noindent Since $\dependencies(m)$ is finite for any $m$, and since the dependency relation is well-founded, the
above identity yields a correct inductive definition for $\psi_\free$.
We then have the following generic validator result for free composition.
\begin{theorem}[\coqref{VLSM.Core.Validators.FreeCompositionValidator}{free_valid_message_is_valid}]\label{thm:free-validator-message}
   For any message $m$, if $\psi_\free(m)$ holds then $m$ is valid in $\mathcal{V}$.
\end{theorem}

\begin{theorem}[\coqref{VLSM.Core.Validators.FreeCompositionValidator}{free_valid_message_yields_projection_validator}]\label{thm:free-validator}
    For any component $\mathcal{V}_i$, if $\beta_i(l_i, s_i, m)$ implies $\psi_\free(m)$ for any $l_i$, $s_i$, $m$, then $\mathcal{V}_i$ is a validator for $\mathcal{V}$.
\end{theorem}

\subsection{Induced validators}

When a component is not a validator for a composition, we can construct a more constrained version of the component which will act as a validator for the composition.

\begin{definition}[Induced validator, \coqref{VLSM.Core.Validator}{composite_vlsm_induced_projection_validator}]\label{induced-validator}
The \textbf{induced  $j^{th}$  validator} of $\mathcal{V}$ is the VLSM
$$\Val_j(\mathcal{V}) = (L_{j}, S_{j}, S_{j,0}, M, M_{j,0},\tau_{j}, (\valid \wedge \varphi)|_j),$$
where $L_{j}, S_{j}, S_{j,0}, M, M_{j,0}, \tau_{j}$ are the same as for the component $\mathcal{V}_j$, while the validity constraint predicate is defined as
\begin{align*}
(\valid \wedge \varphi)|_j \mbox{ holds} \quad  \mbox{iff} & \quad m \mbox{ is valid in } \mathcal{V}  \mbox{ and there is a valid state } \sigma=\langle s_1,\mydots,s_{j-1},s,s_{j+1},\mydots,s_n\rangle  \mbox{ in } \mathcal{V} \\ 
 & \hspace{.3cm}  \mbox{ such that }  (\valid \wedge \varphi)(\langle j,l\rangle,\sigma,m) \mbox{ holds}.
\end{align*}
\end{definition}

The  induced $j^{th}$ validator  for a free composition can be obtained as a particular case of the above definition. 
The $(\valid \wedge \varphi)|_j$ predicate ensures that constrained transitions can be lifted to valid transitions in the composition. 
The induced  $j^{th}$ validator and $\mathcal{V}_j$ do not usually coincide. Even though the induced $j^{th}$ validator has the same states as $\mathcal{V}_j$, it has a potentially different set of valid states than $\mathcal{V}_j$, in particular due to the interactions with other components and the possible composition constraint.

The following lemmas characterise validators induced by validating components.

\begin{lemma}[\coqref{VLSM.Core.Validator}{projection_validator_messages_transitions}]
If $\valid_j(l,s_j,m)$ implies that $(\valid \wedge \varphi)|_j(l,s_j,m)$ holds for any $l \in L_j$,  constrained state $s_j$  in $\mathcal{V}_j$, and $m\in M$,  then $\mathcal{V}_j$ is a validator for $\mathcal{V}$.
\end{lemma}

Similarly with the case of the induced projections, valid traces in the composition under the same constrained composition of the original components and of the induced validators coincide.
The free composition of induced validators does not necessarily globally satisfy $\varphi$, but it is the best approximation of the constrained composition one can obtain using a free composition.

\subsection{Example: VLSMs performing multiplication of inputs}

Let us consider the free composition $\mathcal{P} = \sum_{\mathit{p\ prime}}\mathcal{P}_p$ of the VLSMs $\mathcal{P}_p$ (\coqref{VLSM.Examples.Tutorial.PrimesComposition}{primes_vlsm_composition}) introduced in Subsection \ref{subsec:multiplication}. Note that the valid messages of $\mathcal{P}$ are
$$M^{\mathcal{P}}_{v} = \{2^{k_2} \cdot 3^{k_3} \cdot 5^{k_5} \cdot \mydots\ |\ k_2 + k_3 + k_5 + \mydots  \geq 1\}.$$

\begin{theorem}[\coqref{VLSM.Examples.Tutorial.PrimesComposition}{component_projection_validator_prop_primes}]
For any prime $p$, $\mathcal{P}_p$ is a validator for $\mathcal{P}$.
\end{theorem}

Let us now consider the following composition constraint guaranteeing that the received messages are an even numbers (\coqref{VLSM.Examples.Tutorial.PrimesComposition}{even_constrained_primes_composition})
\begin{align*}
\varphi_{\mathit{parity}} & = \{(\langle \mathcal{P}_2,d\rangle,\langle n_1,n_2,n_3,\mydots \rangle, i)\ |\  i\ \mathrm{mod}\ 2 = 0\}\  \cup \\
 				    & \hspace{.4cm} \{(\langle \mathcal{P}_3,d\rangle,\langle n_1,n_2,n_3,\mydots\rangle, i)\ |\  i\ \mathrm{mod}\ 2 = 0 \} \ \cup \\
				    & \hspace{.4cm} \{(\langle \mathcal{P}_5,d\rangle,\langle n_1,n_2,n_3,\mydots\rangle, i)\ |\ i\ \mathrm{mod}\ 2 = 0 \}  \ \cup \\
				    & \hspace{.4cm} \mydots
\end{align*}
The valid messages in the composition $\mathcal{P}$ constrained by $\varphi_{\mathit{parity}}$ are 
$$M^{\mathcal{P}}_{v} = \{2^{k_2} \cdot 3^{k_3} \cdot 5^{k_5} \cdot \mydots\ |\ k_2 \geq 1\} \cup \{3,5,\ldots\}.$$

For any prime $p$, $\mathcal{P}_p$ is not a validator for $\mathcal{P}$ (\coqref{VLSM.Examples.Tutorial.PrimesComposition}{even_constrained_primes_composition_no_validator}).
However, if we add to any  $\mathcal{P}_p$  the local constraint $$\varphi_{\mathit{parity}}^{\mathcal{P}_p} = \{(d,n, i)\ |\  i\ \mathrm{mod}\ 2 = 0\},$$ then this VLSM becomes a validator for $\mathcal{P}$ (\coqref{VLSM.Examples.Tutorial.PrimesComposition}{even_constrained_primes_composition_all_validators}).

\subsection{Example: Message Observer (MO) components and the MO protocol}\label{MO}

As we noticed at the beginning of the section, in UMO components (Subsection \ref{UMO}) we have no control over the pattern of the received messages and we cannot establish if a message is valid. 
MO components are just refinements of UMO components by strengthening the validity constraint predicate in order to ensure that received messages are valid. 

The \textit{MO component of address $i \in \mathbb{N}$} (\coqref{VLSM.Examples.ELMO.MO}{MO_component}) has the same ingredients as the UMO component of address $i$, except for the validity constraint predicate
$\valid_{\MO}(l,s,m) = \valid_{\UMO}(l,s,m) \wedge (l = \receive \to \psiVal(m))$, where
\begin{center}
	\begin{tabular}{l}
		\hspace{.6cm} $\psiVal(\stateU{[]}{j}) = j \in \{1,\mydots,n\}$, \\[.1cm]
		\hspace{.6cm} $\psiVal(\stateU{o \concat [(l_p,\stateU{o_p}{j_p})]}{j}) =  \psiVal(\stateU{o}{j})\ \wedge$ \\[.1cm] 
		\hspace{1.6cm} $(l_p = \send \rightarrow (j_p = j \wedge o_p = o))\ \wedge  		 
			(l_p = \receive \rightarrow \psiVal(\stateU{o_p}{j_p}))$.
	\end{tabular}
\end{center}

The recursive invocation in formula $\psiVal$ is terminating as the list of message observations contained in a state is finite and it is decreasing at each recursive invocation (\coqref{VLSM.Examples.ELMO.MO}{MO_msg_valid}).

For example, the following trace which was a constrained trace for the UMO component of address $2$, is not a constrained trace for the MO component of address $2$
{
$$
 \stateU{[]}{2}
\xrightarrow[m_1 = \stateU{[]}{2}]{\send}
\stateU{[{(\send, m_1)}]}{2}  
\xdashrightarrow[m_2 = \stateU{[(\send, \stateU{[]}{1}), (\send, \stateU{[]}{2})]}{1} ]{\receive} 
\stateU{[(\send, m_1), (\receive, m_2)]}{2}.
$$
}
as $\psiVal(m_2)$ does not hold.

Let $\{\mathcal{M}_i\}_{i=1}^n$ be a set of MO components indexed by their addresses, i.e, $\mathcal{M}_i$ is the MO component of address $i$. The \textit{MO protocol} $\umo(\mathcal{M}_i)_{i=1}^n$ is defined as the free VLSM composition of $\{\mathcal{M}_i\}_{i=1}^n$ (\coqref{VLSM.Examples.ELMO.MO}{MO}).
Lemmas \ref{UMO-component-trace} and \ref{UMO-protocol-trace} can be proved for MO components and protocols as well (see, e.g., \coqref{VLSM.Examples.ELMO.MO}{constrained_state_contains_unique_constrained_trace}).

Let us show that MO components satisfy the above  generic conditions for validator for free composition. Moreover, $\valid_{\MO}(l,s,m)$ implies $\psi_\free(m)$, for any $l,s,m$.  With this venture, we also show that the theory of VLSMs is in alignment with the correct-by-construction methodology and illustrate how our generic results can be used for particular VLSMs. 

\vspace{-.2cm}
\begin{description}

\item[{\color{gray}\ChannelAuthentication.}] For any message $m$, $\senderM{m} = \id(m)$.

\item[{\color{gray}\MessageDependencies.}] The dependencies of a message $m$ are given by the set of message observations contained in $m$, i.e.,
	$\dependencies(m) = \{m' \mid (l',m') \in \obs(m)\}.$

\item[{\color{gray}\Emittable.}] Since a message contains its full history, we simply replay its transitions. Then the message itself is constrained iff all the replayed transitions are constrained.

\end{description}

\begin{corollary}[\coqref{VLSM.Examples.ELMO.MO}{MO_component_validating}]\label{MO-validator}
Every MO component $\mathcal{M}_i$ is a validator for  $\mo(\mathcal{M}_i)_{i=1}^n$.
\end{corollary}

\section{Evidence of Equivocation}\label{evidence-equivocation}

In the consensus literature, equivocation refers to claiming different beliefs about the state of the protocol to different parts of the network \cite{clement,madsen,jaffe}. Equivocating components behave as-if running multiple copies of the protocol.  For example, if a network is trying to come to consensus about the value of a bit, an equivocating node may claim to think the bit is 0 to one part of the network and 1 to another part.  In CBC Casper, an equivocating component may issue two blocks, neither of which is in the justification of the other \cite{vlad-2019}.  

Pure equivocation is hard to detect, but we can look for {\em evidence of equivocation}. Evidence of equivocation can be either {\em local}, in a single component where we have access only to states of the component, or {\em global}, in a composite system where we have access to composite states.

For example, let us consider two messages for MO components (Subsection \ref{MO}),
$$m_1 = \stateU{[]}{2} \mbox{ and } m_2 = \stateU{[(\receive,m)]}{2}.$$
Consider a state of the MO component of address $1$ in which we have the message observations $$(\receive,m_1) \mbox{ and }  (\receive,m_2).$$
Then $m_1$ and $m_2$ are a local evidence of equivocation for the component of address $2$ as they cannot be emitted by component $2$ in a single run. Indeed, they can only be produced in different traces, e.g.,
\begin{align*}
&\stateU{[]}{2} \xrightarrow[m_1 = \stateU{[]}{2}]{\send} \stateU{[{(\send, m_1)}]}{2} \\
&\stateU{[]}{2} \xrightarrow[m]{\receive} \stateU{[{(\receive, m)}]}{2} \xrightarrow[m_2 =  \stateU{[{(\receive, m)}]}{2}]{\send} \stateU{[{(\receive, m), (\send, m_2)}]}{2}
\end{align*}

\subsection{Local and global evidence of equivocation}
In order to be able to express the notions of local and global evidence of equivocation for a generic VLSM, we need to make the \textit{Channel Authentication assumption} (Assumption \ref{channel-authentication}), the \textit{Message Dependencies assumption} (Assumption \ref{message-dependencies}), and the following two extra assumptions:

\begin{assumption}[Received, \coqref{VLSM.Core.Equivocation}{HasBeenReceivedCapability}]
\label{received}
There exists  an oracle which can tell if a message was received on every trace leading to a component state. For any message $m$ and state $s$, we denote this oracle predicate by $\receivedM{s}{m}$.
It guarantees that any trace reaching the component state would contain this message. This assumption can be naturally lifted to composite states (\coqref{VLSM.Core.Equivocation}{free_composite_HasBeenReceivedCapability}). 

\end{assumption}

\begin{assumption}[Sent, \coqref{VLSM.Core.Equivocation}{HasBeenSentCapability}]
\label{sent}
There exists an oracle which can tell if a message was sent on every trace leading to a component state. For any message $m$ and state $s$, we denote this oracle by $\sentM{s}{m}$.  This assumption guarantees that any trace reaching the component state would contain this message. This assumption can be naturally lifted to composite states (\coqref{VLSM.Core.Equivocation}{free_composite_HasBeenSentCapability}). 
\end{assumption}

\begin{figure}[ht]
\begin{tikzpicture}

\node[] (msgDep) at (-1,0) {\footnotesize \shortstack{message dependencies\\assumption}};
\node[] (msgDepInd) at (-1,-1.25) {\footnotesize \shortstack{indirect message \\ dependency relation}};
\node[] (beenDirObs) at (-1,-2.5) {\footnotesize \shortstack{directly observed\\information}};

\node[] (beenSent) at (-4.5,-3) {\footnotesize \shortstack{sent\\assumption}};
\node[] (beenReceived) at (-4.5,-2) {\footnotesize \shortstack{received\\assumption}};

\node[] (channelAuth) at (3,0) {\footnotesize \shortstack{channel authentication\\assumption}};
\node[] (inc) at (3,-1.25) {\footnotesize \shortstack{incomparable\\messages}};
\node[] (beenObs) at (3,-2.5) {\footnotesize \shortstack{indirectly observed\\information}};

\node[] (localEquiv) at (6,-1.25) {\footnotesize \shortstack{local \\equivocation}};

\path[->, line width=1pt, mDarkTeal!50] (msgDep) edge (msgDepInd);
\path[->, line width=1pt, mDarkTeal!50] (msgDepInd) edge (inc);
\path[->, line width=1pt, mDarkTeal!50] (channelAuth) edge (inc);
\path[->, line width=1pt, mDarkTeal!50] (beenSent) edge (beenDirObs);
\path[->, line width=1pt, mDarkTeal!50] (beenReceived) edge (beenDirObs);
\path[->, line width=1pt, mDarkTeal!50] (msgDepInd) edge (beenObs);
\path[->, line width=1pt, mDarkTeal!50] (beenDirObs) edge (beenObs);
\path[->, line width=1pt, mDarkTeal!50] (channelAuth) edge (localEquiv);
\path[->, line width=1pt, mDarkTeal!50] (inc) edge (localEquiv);
\path[->, line width=1pt, mDarkTeal!50] (beenObs) edge (localEquiv);
\end{tikzpicture}
\centering
\caption{Assumptions and derived notions needed for local evidence of equivocation. Nodes with no incoming arrows represent assumptions, while the rest represent derived notions. The arrows show the dependencies between them.} 
\label{fig:local-equivocation}
\end{figure}
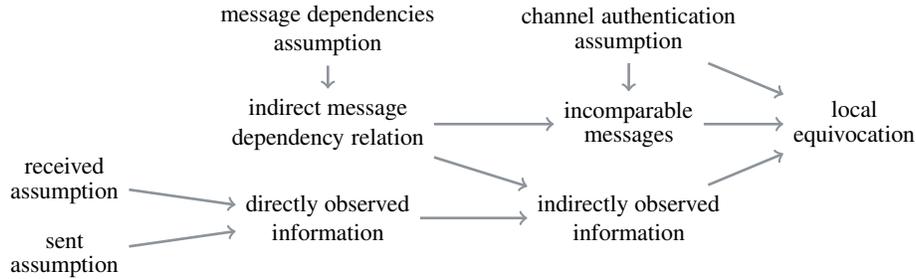

\begin{figure}[t]
\begin{tikzpicture}

\node[] (msgDep) at (-1,0) {\footnotesize\shortstack{message dependencies\\assumption}};
\node[] (msgDepInd) at (-1,-1.25) {\footnotesize\shortstack{indirect message \\ dependency relation}};
\node[] (beenDirObs) at (-1,-2.5) {\footnotesize\textbf{\shortstack{directly observed\\information}}};

\node[] (beenSent) at (-4.5,-3) {\footnotesize\textbf{\shortstack{emitted\\assumption}}};
\node[] (beenReceived) at (-4.5,-2) {\footnotesize\textbf{\shortstack{received\\assumption}}};

\node[] (channelAuth) at (3,0) {\footnotesize\shortstack{channel authentification\\assumption}};
\node[] (beenObs) at (3,-2.5) {\footnotesize\textbf{\shortstack{indirectly observed\\information}}};

\node[] (globalEquiv) at (6,-1.25) {\footnotesize\textbf{\shortstack{global \\equivocation}}};

\path[->, line width=1pt, mDarkTeal!50] (msgDep) edge (msgDepInd);
\path[->, line width=1pt, mDarkTeal!50] (channelAuth) edge (globalEquiv);
\path[->, line width=1pt, mDarkTeal!50] (beenSent) edge (beenDirObs);
\path[->, line width=1pt, mDarkTeal!50] (beenReceived) edge (beenDirObs);
\path[->, line width=1pt, mDarkTeal!50] (msgDepInd) edge (beenObs);
\path[->, line width=1pt, mDarkTeal!50] (beenDirObs) edge (beenObs);
\path[->, line width=1pt, mDarkTeal!50] (beenObs) edge (globalEquiv);
\path[-, line width=1pt, mDarkTeal!50] (beenSent) edge (-4.5,-3.7);
\path[-, line width=1pt, mDarkTeal!50] (-4.5,-3.7) edge (6,-3.7);
\path[->, line width=1pt, mDarkTeal!50] (6,-3.7) edge (globalEquiv);
\end{tikzpicture}
\centering
\caption{Assumptions and derived notions needed for global evidence of equivocation. The nodes depicted in boldface are for composite states.}
\label{fig:global-equivocation}
\end{figure}
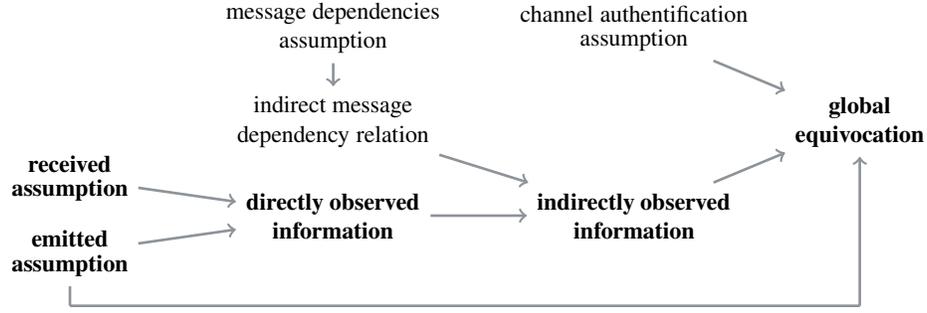

From the above assumptions, we can derive the following notions needed to formalise the concepts of local and global equivocation. We illustrate the connections between them in Figures \ref{fig:local-equivocation} and \ref{fig:global-equivocation}.

\begin{description}
\item [\IndirectMessageDependencies] can be extracted as the transitive closure of the dependency relation between messages implied by the \messageDependencies
\ (\coqref{VLSM.Core.MessageDependencies}{msg_dep_happens_before}). 

\item [\IncomparableMessages] is a way of expressing when two messages which have the same sender are incomparable with respect to the indirect message dependency relation.

\item [\HasBeenDirectlyObserved] is a way of expressing if a message is directly observed in a component state, using the received and emitted assumptions
(\coqref{VLSM.Core.Equivocation}{HasBeenDirectlyObservedCapability}, \coqref{VLSM.Core.Equivocation}{HasBeenDirectlyObservedCapability_from_sent_received}). 
This notion can be naturally lifted to composite states (\coqref{VLSM.Core.Equivocation}{free_composite_HasBeenDirectlyObservedCapability}).

\item [\HasBeenObserved.] A message is indirectly observed in a component state if either it is directly observed in the state, or it is an indirect dependency of a directly observed message in the state (\coqref{VLSM.Core.MessageDependencies}{HasBeenObserved}). 
This notion can be naturally lifted to composite states (\coqref{VLSM.Core.MessageDependencies}{composite_HasBeenObserved_iff}).
\end{description}

Now we can define the notions of evidence of equivocation mentioned at the beginning of this section.

\begin{definition}[Local evidence of equivocation, \coqref{VLSM.Core.MessageDependencies}{msg_dep_is_locally_equivocating}]\label{local-equivocation}
A pair of messages is a \textbf{local evidence of equivocation} for their sender in a component state of a VLSM if
\begin{list}
{($\arabic{cont}$)}{\usecounter{cont}}
	\item the messages have the same sender,
	\item the messages have been (indirectly) observed in the component state, and
	\item the messages could not have been produced by their sender in a single run of the protocol.
\end{list}
We denote by $\localEquiv(s)$ the set of components for which there is a local evidence of equivocation in a component state $s$.
\end{definition}

\begin{definition}[Global evidence of equivocation, \coqref{VLSM.Core.MessageDependencies}{msg_dep_is_globally_equivocating}]\label{global-equivocation}
A message is a \textbf{global evidence of equivocation} for its sender  in a composite state of a composite VLSM if
\begin{list}
{($\arabic{cont}$)}{\usecounter{cont}}
	\item the message has been (indirectly) observed in the composite state, and 
	\item the message was not observed as a sent message in the composite state. 
\end{list}
We denote by $\globalEquiv(\sigma)$ the set of components for which there is a global evidence of equivocation in a composite state $\sigma$.
\end{definition}

Let us now investigate some properties of the above evidences of equivocation notions.

\begin{theorem}[\coqref{VLSM.Core.MessageDependencies}{msg_dep_locally_is_globally_equivocating}]
Let $\mathcal{V} =\ \bigr({\sum_{i=1}^n} \mathcal{V}_i\bigr)\bigr|_\varphi $ be the constrained composition under $\varphi$ of $\{\mathcal{V}_i\}_{i=1}^n$.  For any constrained component state $s$ and any constrained composite state $\sigma$ such that one of its components is $s$, we have
$$\localEquiv(s) \subseteq \globalEquiv(\sigma).$$
\end{theorem}

Note that the other implication is not true in general. The local evidence of equivocation exposed by a component is persistent, in the sense that whatever state the protocol transitions to, this equivocation will still be exposed. In contrast,  the global evidence of equivocation is not persistent as, for example, the message that was an evidence of equivocation in a composite state $\sigma$ could be observed as a sent message in a composite state $\sigma'$ reachable from $\sigma$. However, we can construct a trace reaching a composite state which does not expose more global evidence of equivocation at any step than that composite state.

\begin{theorem}[\coqref{VLSM.Core.Equivocation.MinimalEquivocationTrace}{state_to_minimal_equivocation_trace_equivocation_monotonic}]
Let $\{\mathcal{V}_i\}_{i=1}^n$ be an indexed set of VLSMs such that for each component the state successor relation induced by its transition function is well-founded.
Let $\mathcal{V} =\ \bigr({\sum_{i=1}^n} \mathcal{V}_i\bigr)\bigr|_\varphi $ be the constrained composition under  $\varphi$ of $\{\mathcal{V}_i\}_{i=1}^n$ and $\sigma$ be a constrained composite state of $\mathcal{V}$. Then there is a constrained trace reaching $\sigma$ such that for any composite states $\sigma'$ and $\sigma''$ in this trace, $\sigma'$ appearing before $\sigma''$, we have
$$\globalEquiv(\sigma') \subseteq \globalEquiv(\sigma'') \subseteq \globalEquiv(\sigma).$$ 
\end{theorem}

\subsection{Example: Evidence of equivocations in MO components and the MO protocol}

We show that MO components and MO protocols (Subsection \ref{MO}) satisfy the above assumptions needed for the notions of  evidence of equivocation, hence obtaining the notions of local and global evidence of equivocation for MO components derived from Definitions \ref{local-equivocation} and \ref{global-equivocation}.
In Subsection \ref{MO}, we already showed that MO components satisfy the \ChannelAuthentication\ and the \MessageDependencies.

\vspace{-.2cm}
\begin{description}

\item [{\color{gray}\Received.}] A state $s$ has received a message $m$ if  $(\receive,m) \in \obs(s)$. 

\item [{\color{gray}\Sent.}] A state $s$ has emitted a message $m$ if $(\send,m) \in \obs(s)$. 
\item [{\color{gray}\IndirectMessageDependencies.}] It is the least transitive relation, denoted by $<$, satisfying the condition $m_1 < m_2$ whenever $m_1 \in \dependencies (m_2)$. As a particular case, we say that two messages $m_1$ and $m_2$ are \textit{sent-comparable} if in $m_2$ we can observe that $m_1$ was already emitted (i.e., $m_1 \in \{m' \mid (\send,m') \in \obs(m_2))$.

\item [{\color{gray}\IncomparableMessages.}] We say that two messages $m_1$ and $m_2$  are incomparable, denoted by $m_1 \incomp m_2$, if they have the same sender and they are not sent-comparable.

\item [{\color{gray}\HasBeenDirectlyObserved.}] A  message $m$ is directly observed in $s$ if $m \in \messages(s)$.

\item [{\color{gray}\HasBeenObserved.}] A message $m$ is observed in a state $s$ if $m < s$, where $s$ is viewed as a message in this instance. 

\end{description}

\subsection{Limited equivocation}

Most consensus protocols require some bound on the number of equivocating parties. This can be achieved by means of a composition constraint which keeps the global evidence of equivocation under a threshold. However, the components do not have access to global information and therefore can keep track only of local evidence of equivocation. Some generalize away from the number of parties to the $\mathit{weight}$ of the parties; this reduces to the count when the weights are all $1$.  Therefore 
we assume additionally that there is a \textit{weight} function from the (addresses of) components to positive real numbers and a fixed {\em equivocator threshold} $t$. 

We can make one further assumption about the VLSMs involved which can simplify the process of detecting equivocation. 

\begin{assumption}[Full node, \coqref{VLSM.Core.MessageDependencies}{message_dependencies_full_node_condition_prop}]
\label{full-node}
There exists an oracle which ensures that before receiving a message, a VLSM has previously observed all the message dependencies of that message.
\end{assumption}

The {\em \fullNode} is a way of limiting the amount of new equivocation when receiving a message.  Under the full node assumption, the only new equivocation which can be introduced when receiving a message is that of the sender of the message, since the equivocation introduced by its dependencies has already been accounted for. 
In this setting, the notions of evidences of equivocation can be further simplified by only considering directly observed messages in a state. 
We illustrate this scenario in the example from Subsection \ref{ELMO}.

\subsection{Example: Limited equivocation in ELMO components and the ELMO protocol}\label{ELMO}

When dealing with equivocation, a design goal is to limit the global equivocation. This can be achieved by means of a composition constraint which keeps the global evidence of equivocation under a threshold. However, the components do not have access to global information and therefore can keep track only of local evidence of equivocation.

 We will illustrate these ideas with a refinement of MO components called ELMO components. ELMO is an acronym for {\em Equivocation-Limited Message Observer}. We will define an ELMO protocol as a constrained composition of ELMO components which will ensure that the  global equivocation exhibited by the system remains under a fixed threshold. We will show that ELMO components are validators for an ELMO protocol, meaning that they are able to locally impose this condition on equivocation.

Let us fix an  {\em equivocation threshold} $t$ (a positive real number) and a function $\weight$ from component addresses  to positive real numbers. The function $\weight$ will be used to count the weight of the equivocating components; this reduces to the count when the weights are all $1$.

In order to define ELMO components, we need to introduce some extra definitions for MO components. Let $\mathcal{M}_i$ be the MO component of address $i$.
The following  is a refinement of the notion of local evidence of equivocation under the full node assumption and relies on the assumption that a component cannot self-equivocate. 
Let $s$ be a state of $\mathcal{M}_i$ with $\obs(s) = \obs(s') \concat [(l,m)]$. The set of \textit{locally evidenced equivocators under full node assumption} ([\coqref{VLSM.Examples.ELMO.ELMO}{local_equivocators_full}]) in $s$ is:
\begin{align*}
\localEquivocatorsFull(s) =
\left \{
\begin{array}{rl}
 \localEquivocatorsFull(s') \cup \{\id(m)\}, & \mbox{if } l = \receive \mbox{ and there exists } \\
 &   m' \in \receivedmessages(s')
    \mbox{ such that } m \incomp m',  \\
 \localEquivocatorsFull(s'), & \mbox{otherwise.}
\end{array}
\right. 
\end{align*}
    
The \textit{ELMO component of address $i \in \mathbb{N}$}  (\coqref{VLSM.Examples.ELMO.ELMO}{ELMO_component_machine}) has the same ingredients as the MO component of address $i$, except for the validity constraint predicate
$$\valid_{\ELMO}(l,s,m) = \valid_{\UMO}(l,s,m)\ \wedge\ (l = \receive \rightarrow \psi_{\ELMO}(s,m)),$$

where $\psi_{\ELMO}(s,m)$ is the conjunction of the following predicates:
\begin{align*}	
&\psi_{\mathit{fullNode}}(s,m) = \dependencies(m) \subseteq \messages(s), \\
&\psi_{\mathit{noSelfEqv}}(s,m) = (\id(m) = i\, \wedge\, \id(s) = i) \rightarrow m \in \sentmessages(s), \\
&\psi_{\mathit{msgValidFull}}(\state{[]}{j}) = j \in \{1,\mydots,n\}, \\
&\psi_{\mathit{msgValidFull}}(\state{o \concat [(l_p,\state{o_p}{j_p})]}{j}) =  \psi_{\mathit{msgValidFull}}(\state{o}{j})\ \wedge \\
			 &\hspace{1cm} (l_p = \send \rightarrow j_p = j \wedge o_p = o)\ \wedge\  \\
		 	 &\hspace{1.5cm} (l_p = \receive \rightarrow \psi_{\mathit{fullNode}}(\state{o}{j},\state{o_p}{j_p}) \wedge \psi_{\mathit{noSelfEqv}}(\state{o}{j},\state{o_p}{j_p})), \\
&\psi_{eqv}(s) = \sum_{j \in \localEquivocatorsFull(s)}\weight(j) < t.
\end{align*}

The validity constraint for an ELMO component enforces the full node assumption, checks message validity, ensures that the component does not self-equivocate, and only allows receiving  a message if this action will not bring the total weight of locally-visible equivocating components above the equivocation threshold.
For ELMO components, we can prove easily that the two notions of local evidence of equivocation coincide.

\begin{lemma}[\coqref{VLSM.Examples.ELMO.ELMO}{local_equivocators_iff_full}]
Let $\mathcal{E}_i$ be the ELMO component of address $i$. For any constrained state $s$  of $\mathcal{E}_i$, we have 
$$\localEquiv(s) = \localEquivocatorsFull(s).$$
\end{lemma}

Let us denote by $\globalEquivocatorsFull$ the notion of global evidence of equivocation under the full node assumption. 
Let $\{\mathcal{E}_i\}_{i=1}^n$ be a set of ELMO components indexed by their addresses, i.e, $\mathcal{E}_i$ is the ELMO component of address $i$. The \textit{ELMO protocol} $\elmo(\mathcal{E}_i)_{i=1}^n$ is defined as the constrained composition of $\{\mathcal{E}_i\}_{i=1}^n$ under the composition constraint
$$\varphi_{\elmo}(\langle i,\receive \rangle, \sigma , m)  = \sum_{j\ \in\ \globalEquivocatorsFull(\sigma')} \weight (j) < t,$$
where $\sigma' = \tau^s(\langle i,\receive \rangle, \sigma , m)$.

The following result shows that the local validity constraints of the ELMO components are strong enough to ensure that the  composition constraint holds.

\begin{theorem}[\coqref{VLSM.Examples.ELMO.ELMO}{ELMO_components_validating}]
Let $\mathcal{E} = \elmo(\mathcal{E}_i)_{i=1}^n$ be an ELMO protocol. Every component $\mathcal{E}_i$  is a validator for $\mathcal{E}$.
\end{theorem}

\section{Models of Equivocation}\label{modelsOfEquivocation}

As explained in the previous section, equivocation occurs when receiving a message which has not been previously sent in the current trace;  the sender of the message is then said to be equivocating. In this section we introduce two models of equivocation in the VLSM framework: the \textit{state equivocation model} and the \textit{message equivocation model}. These models allow one to model equivocation within a (composed) VLSM, as their transition functions and validity constraints formalise the idea of evidence of equivocation internally in the (composed) VLSM.  We investigate when  these two models of equivocation are equivalent. We consider two scenarios: (1) a fixed subset of components can equivocate, and (2) the set of equivocating components is weight-limited.

Let $\{\mathcal{V}_i\}_{i=1}^n$  be an indexed set of VLSMs over the same set of messages. We assume that each $\mathcal{V}_i$ satisfies the {\em  \channelAuthentication} (Assumption \ref{channel-authentication}), the  {\em \sent\  } (Assumption \ref{sent}), and the \textit{ full node assumption} (Assumption \ref{full-node}).
\subsection{State and message equivocation models for a fixed set of equivocators}

We fix a subset $E\subseteq \{1,\mydots,n\}$.  We assume that the only components which can equivocate are those $\mathcal{V}_i$ with $i \in E$. Let $\mathit{\mathit{NonE}}$ denote the non-equivocating components, i.e.,  $\mathit{\mathit{NonE}} = \{1,\mydots,n\}\setminus E$. 

Below we describe the two models of equivocation for this scenario and investigate the conditions under which they are equivalent.

\subsubsection{The state equivocation model}
%
 In this model, we allow an equivocating component to perform \textit{state equivocations} by forking itself or spawning new machines. In order to model this, we associate to any VLSM its {\em equivocating version} which can fork itself or spawn new copies of itself at any moment. A state for the equivocating version of an VLSM is a list of states, each element of the list being a state of the original VLSM; the state of the equivocating VLSM keeps track of all the possible states that can be reached by equivocation. A VLSM and its equivocating version are defined over the same set of messages. Formally, we have the following definition.

\begin{definition}[Fixed state equivocation model, \coqref{VLSM.Core.Equivocators.Equivocators}{equivocator_vlsm}]\label{equivocator-VLSM}
Let $\mathcal{V} = (L,S, S_{0}, M, M_0, \tau, \valid)$ be a VLSM. The \textbf{equivocating VLSM} associated with $\mathcal{V}$ is the VLSM  $$\mathcal{V}^e = (L^e,S^e, S_{0}^e, M, M_{0}, \tau^e, \valid^e),$$ 
where $L^e = \mathbb{N}^* \times (L \cup \{\mathrm{duplicate}\} \cup (\{\mathrm{new\_machine}\} \times S_{0}))$, $S^e = [S]$\footnote{For any set $A$, we denote by $[A]$ the set of all finite lists over $A$. For any list $l$ over $A$, we denote by $\len(l)$ the length of $l$ and by $l[n]$ its nth element.}, $S_{0}^e = \{[s]\ |\ s \in S_0\}$, the transition function is defined as
{\small
\begin{align*}
&\tau^e(\langle i, l \rangle, \gamma,m)  = ([\gamma[1],\mydots,\gamma[i-1],\tau^{s}(l,\gamma[i],m),\gamma[i+1],\mydots], \tau^{m}(l,\gamma[i],m)) \mbox{ with } l \in L,\\
&\tau^e(\langle i, \mathrm{duplicate} \rangle, \gamma,m)  = ([\gamma[1],\mydots,\gamma[i-1],\gamma[i],\gamma[i],\gamma[i+1],\mydots], \nomessage),\\
&\tau^e(\langle i, (\mathrm{new\_machine},s_{0}) \rangle, \gamma,m)  = ([\gamma[1],\mydots,\gamma[i],s_{0},\gamma[i+1],\mydots], \nomessage),
\end{align*}
}
while the validity constraint is defined as
{\small
\begin{align*}
&\valid^e(\langle i, l \rangle,\gamma,m) = i \leq \len(\gamma) \wedge \valid(l,\gamma[i],m), \hspace{4cm} \\
&\valid^e(\langle i, \mathrm{duplicate}\rangle, \gamma ,\nomessage) = i \leq \len(\gamma), \\ 
&\valid^e(\langle i, (\mathrm{new\_machine},s_{0}) \rangle, \gamma ,\nomessage) = i \leq \len(\gamma).
\end{align*}
}
\end{definition}

For an equivocating VLSM, the first component of a label indicates which copy of the machine will be used for a transition, while the validity constraint  ensures that we can refer to an already existing copy of the component.  
We can show that since $\mathcal{V}$ satisfies the {\em \sent} (Assumption \ref{sent}), then $\mathcal{V}^e$ also satisfies the {\em \sent\ } by considering that  a message was emitted on any trace leading to a state of the equivocating VLSM if there is a copy of the VLSM in that state for which the message  was emitted on any trace leading to it (\coqref{VLSM.Core.Equivocators.MessageProperties}{equivocator_HasBeenSentCapability}).

We can now define the {\em state equivocation model for a fixed set of equivocators $E$} as a VLSM composition in which we impose the composition constraint that components may only receive messages that have been sent in the current trace of the composition.

\begin{definition}[Fixed state equivocation model, \coqref{VLSM.Core.Equivocators.FixedEquivocation}{equivocators_fixed_equivocations_vlsm}]
The \textbf{state equivocation model of $\{\mathcal{V}_i\}_{i=1}^n$ for the fixed set of equivocators $E$} is the constrained composition in which we replace each component which can equivocate by its corresponding equivocating VLSM and use the composition constraint that components may only receive messages that have been sent in the current trace. Formally, we have
$$\mathcal{V}_{s\_\Eqv}^E =\ \Bigr({\sum_{i=1}^n} \mathcal{V}_i' \Bigr) \Bigr|_{\varphi_{s\_\Eqv}} = (L,S, S_0, M, M_0,\tau, \valid \wedge \varphi_{s\_\Eqv}) $$
where, for any $1\leq i \leq n$, $\mathcal{V}_i' = \mathcal{V}_i$ if $i\not\in E$ and $\mathcal{V}_i' = \mathcal{V}_i^e$ if $i\in E$, and
\[
\varphi_{s\_\Eqv}( \iota , \langle \gamma_1,\mydots,\gamma_n\rangle ,m) = \sentM{\gamma_{\senderM{m}}}{m}.
\]
\end{definition}

At any point, each equivocating component can perform a state equivocation either by making a copy of one of the states or introducing a new initial state.
Each copy of an equivocating component can evolve independently, but can only receive messages that appear in the current trace of this new machine. 

Given a state of the form $\gamma = \langle \gamma_1,\mydots,\gamma_n \rangle$ in $\mathcal{V}_{s\_\Eqv}^E$, its {\em state reduct} to a state of the composition of $\{\mathcal{V}_i\}_{i=1}^n$ is of the form $\overline{\gamma} = \langle s_1,\mydots,s_n\rangle$, where $s_i$ is $\gamma_i$ if $i \not\in E$ and $s_i$ is $\gamma_i${\footnotesize$[1]$} otherwise.
 Given a transition in $\mathcal{V}_{s\_\Eqv}^E$ of the form
$\transition{\langle j,\langle 1, l_j\rangle \rangle}{\gamma}{m}{\gamma'}{m'},$
its {\em transition reduct} to a transition in the composition of $\{\mathcal{V}_i\}_{i=1}^n$ is 
$\transition{\langle j, l_j\rangle}{\overline{\gamma}}{m}{\overline{\gamma'}}{m'}.$
Given a trace $\tr$ in $\mathcal{V}_{s\_\Eqv}^E$, its {\em trace reduct} to a trace of the composition of $\{\mathcal{V}_i\}_{i=1}^n$ consists of the transition reducts of all transitions with labels of the form $\langle j,\langle 1, l_j\rangle \rangle$, taken in the same order as  in $\tr$.


\begin{example}\label{state-model-example}
{\normalfont
Let us consider the MO components $\mathcal{M}_1$ and $\mathcal{M}_2$ of addresses $1$ and $2$, respectively. We assume that $E = \{2\}$, meaning that only $\mathcal{M}_2$ can equivocate. The following is a constrained trace of the state-equivocation model of $\mathcal{M}_1$ and $\mathcal{M}_2$ for the fixed-set of equivocators $E$:

\begin{align*}
&\begin{bmatrix}
\state{[]}{1} \\
[\state{[]}{2}] \\
\end{bmatrix}
\xrightarrow[m_1 = \state{[]}{1}]{\langle 1, \send \rangle}
\begin{bmatrix}
\state{[(\send,m_1)]}{1} \\
[\state{[]}{2}] \\
\end{bmatrix}
\xrightarrow[m_2 = \state{[]}{2}]{\langle 2, \langle 1, \send \rangle \rangle}
\begin{bmatrix}
\state{[(\send,m_1)]}{1} \\
[\state{[(\send,m_2)]}{2}] \\
\end{bmatrix}\\[1em]
&\xrightarrow[\nomessage]{\langle 2, \langle 1, (\mathrm{new\_machine},\state{[]}{2})\rangle \rangle}
\begin{bmatrix}
\state{[(\send,m_1)]}{1} \\
[\state{[(\send,m_2)]}{2}, \state{[]}{2}] \\
\end{bmatrix}\\[1em]
&\xrightarrow[m_1]{\langle 2, \langle 2, \receive \rangle \rangle}
\begin{bmatrix}
\state{[(\send,m_1)]}{1} \\
[\state{[(\send,m_2)]}{2}, \state{[(\receive,m_1)]}{2}] \\
\end{bmatrix}\\[1em]
&\xrightarrow[m_3 = \state{[(\receive,m_1)]}{2}]{\langle 2, \langle 2, \send \rangle \rangle}
\begin{bmatrix}
\state{[(\send,m_1)]}{1} \\
[\state{[(\send,m_2)]}{2}, \state{[(\receive,m_1),(\send,m_3)]}{2}] \\
\end{bmatrix}
\end{align*}

If after these transitions, $\mathcal{M}_1$ would receive the messages $m_2$ and $m_3$ emitted by $\mathcal{M}_2$ then it could infer that $m_2$ and $m_3$ are a local evidence that $\mathcal{M}_2$ is equivocating  in the sense described in Section \ref{evidence-equivocation}.
}
\end{example}

For the empty set of equivocators, the state equivocation model coincides with the VLSM composition under the composition constraint which ensures that components may only receive messages that have been sent in a current trace of the composition. On the other hand, if we take the set of equivocators to be the whole set of indices, we obtain a composition in which each component can state equivocate freely, while message equivocation is still not allowed.

\subsubsection{The message equivocation model}\label{fix-set message-equivocation}

In the VLSM framework, messages are always available for receiving, even if they were not emitted or it might not be valid to receive them. As explained previously, a \textit{message equivocation} is the receipt of a message that has not yet been sent in that trace. We  add a composition constraint which ensures that only the components in $E$ can exhibit equivocating behavior.

\begin{definition}[Fixed message equivocation model, \coqref{VLSM.Core.Equivocation.MsgDepFixedSetEquivocation}{full_node_fixed_set_equivocation}]\label{message-equivocation-I}
The \textbf{message equivocation model of $\{\mathcal{V}_i\}_{i=1}^n$ for the fixed set of equivocators $E$}  is the constrained composition
\begin{align*}
\mathcal{V}_{m\_\Eqv}^E =\ \Bigr({\sum_{i=1}^n} \mathcal{V}_i \Bigr) \Bigr|_{\varphi_{m\_\Eqv}} &= (L,S, S_0, M, M_0,\tau, \valid \wedge {\varphi_{\Eqv}}), \mbox{ where} \\
\varphi_{m\_\Eqv}(\iota, \sigma,m) &= \sentM{s_{\senderM{m}}}{m} \vee \senderM{m} \in E.
\end{align*}
\end{definition}

We call the (projected) traces in $\mathit{Tr}(\mathcal{V}_{\Eqv}^E)_{\mathit{NonE}}$ {\em traces exposed to $E$-fixed equivocation behavior} (See Section \ref{composition:constrained} for more details on projected traces).


We can consider the \textit{\noMessageEquivocation}\   to be the composition constraint which ensures that components may only receive messages that have been sent in the current trace of the composition. This assumption depends on the {\em \sent}\  and {\em \channelAuthentication}. Formally, the \noMessageEquivocation\  means is defined as
$$
\varphi_{no\_equiv}(\langle i,l\rangle, \langle s_1,\mydots,s_n\rangle,m) = \sentM{s_{\senderM{m}}}{m}.$$

Note that for the empty set of equivocators, the above message equivocation model coincides with the VLSM composition under the \noMessageEquivocation.
On the other hand, if we take the set of equivocators to be the whole set of indices, then the composition constraint always holds, so $\mathcal{V}_{m\_\Eqv}^{\{1,\mydots,n\}}$ and ${\sum_{i=1}^n} \mathcal{V}_i$ coincide.

\subsubsection{Equivalence between the state and message equivocation models.}

Let us investigate when the two models coincide.  We begin by analysing the following example.
 
\begin{example}\label{singleton-equivocation}
{\normalfont
Let us consider the state and message equivocation models for just one VLSM $\mathcal{V}$ which is also an equivocator. Observe that $\mathcal{V}_{s\_\Eqv}^E$ and $\mathcal{V}_{m\_\Eqv}^E$ do not coincide. However, consider only the first state in the list of states maintained by $\mathcal{V}_{s\_\Eqv}^E$.  Even though a copy of $\mathcal{V}$ in the list cannot receive a message unless at least one of the copies has sent the message earlier in the trace, $\mathcal{V}_{s\_\Eqv}^E$ has the ability to copy the initial state, run that copy ahead to the point where a message has been produced, then go back to the first copy in the list and have that one receive the message.  Therefore, given a constrained trace of $\mathcal{V}_{s\_\Eqv}^E$, the subtrace consisting of all the constrained transitions using only labels of the form $\langle 1,l \rangle$, with $l$ a label in $\mathcal{V}$, is a constrained trace of $\mathcal{V}_{m\_\Eqv}^E$, and all constrained traces of $\mathcal{V}_{m\_\Eqv}^E$ arise in this way.  It is in this sense that we consider message-equivocation to be equivalent to state-equivocation.
}
\end{example}

For any indexed family of VLSMs, we might hope that the traces are the same in the two models for equivocation if we restrict our attention to the first element of each list maintained by a component in $E$, as happened in Example \ref{singleton-equivocation}.  This, however, is not true.  While components can receive messages from any trace of a message equivocator, state equivocators rely on interacting with other components.  If those other components do not equivocate, it restricts the state equivocators' behaviour in the future.  For example, suppose that we have three components $\mathcal{V}_0, \mathcal{V}_1$, and $\mathcal{V}_2,$ where $\mathcal{V}_1$ is an equivocator.  $\mathcal{V}_0$ can send either the message $a$ or the message $b,$ but not both.  $\mathcal{V}_1$ sends $c$ in response to $a$ or $d$ in response to $b$.  In the message-equivocation model, $\mathcal{V}_2$ can receive both $c$ and $d$, but in the state-equivocation model, it cannot.

Since an equivocating VLSM is a collection of copies of the original VLSM, it is clear that if the original VLSM satisfies the \fullNode\ then each of the copies satisfies the full node assumption.
Assuming each component from an indexed family of VLSMs satisfies the \fullNode, then their state-equivocation model also satisfies the \fullNode, as it does not introduce new messages.

We can extend this remark to VLSMs with a composition constraint $\varphi$ by checking if there is some valid state among the states in the product of the equivocator states.  However, under some simple composition constraints, state-equivocation is not equivalent to message-equivocation as can be seen in the next example.

\begin{example}
{\normalfont
Consider two VLSMs $\mathcal{V}_0$ and $\mathcal{V}_1$, $\mathcal{V}_0$ being an equivocator. The component $\mathcal{V}_0$ has states $s_0, s_1$, $s_0$ being initial, while $\mathcal{V}_1$ has states $q_0, q_1$, $q_0$ being initial.  There is a single message $m$ and three labels, $l_0, l_1$ and $l_2$.  The initial state of the composition is $(s_0, q_0).$  Let us consider a composition constraint which allows the composite VLSM to transition under $l_0$ to $(s_1, q_0)$ or under $l_1$ to $(s_0, q_1)$, while the state $(s_1, q_1)$ transitions under $l_2$ to itself and sends the message $m$. In the message-equivocation model, the state $(s_1, q_1)$ is unreachable and $m$ is not valid.  But in the state-equivocation model, the system can evolve as follows and emit $m$:
\begin{align*}
\langle [s_0],q_0  \rangle\xrightarrow[\nomessage\ \rightarrow\ \nomessage]{\langle 1, (1,l_0)\rangle} 
\langle [s_1],q_0 \rangle \xrightarrow[\nomessage\ \rightarrow\ \nomessage]{\langle 1,(1,(\mathrm{new\_machine},s_0))\rangle}
\langle [s_0,s_1], q_0 \rangle 
\xrightarrow[\nomessage\ \rightarrow\ \nomessage]{\langle 2, l_1\rangle} \langle [s_0,s_1], q_1 \rangle
\xrightarrow[\nomessage\ \rightarrow\ m]{\langle 1, (2,l_2)\rangle}
\langle [s_0,s_1], q_1 \rangle.
\end{align*}
}
\end{example}

Therefore, for message and state-equivocation to be equivalent, we cannot support all composition constraints.  However, we can support free compositions (as a special case of fixed-set equivocation). We will also examine the case of weight-limited equivocation in the next section.

\begin{theorem}\label{fixed-set-equivocation}
For any fixed-set of equivocators, 
\begin{enumerate}
    \item The trace reduct of a valid trace of the state equivocation model is a valid trace for the message equivocation model (\coqref{VLSM.Core.Equivocators.FixedEquivocation}{fixed_equivocators_vlsm_projection}).
    \item Under \fullNode\  for each component, each valid trace for the message equivocation model can be ``lifted'' to a valid trace for the state equivocation model such that its trace reduct  is the original trace (\coqref{VLSM.Core.Equivocators.FixedEquivocationSimulation}{fixed_equivocators_finite_valid_trace_init_to_rev}).
\end{enumerate}
\end{theorem}
\subsection{State and message equivocation models for a weight-limited set of equivocators}

Most consensus protocols require some bound on the number of equivocating parties. Some generalise away from the number of parties to the $\mathit{weight}$ of the parties; this reduces to the count when the weights are all 1. In addition with the initial assumptions, we assume that the indexed set of VLSMs  $\{\mathcal{V}_i\}_{i=1}^n$ is equipped with a function $\weight$ from the (addresses of) components to positive real numbers. 

Let us fix an {\em equivocator threshold} $t$. We describe below the two models of equivocation for this scenario and investigate again when they are equivalent.

\subsubsection{The state equivocation model}

In this scenario, we allow all VLSMs to state equivocate but place a limit on the total weight of equivocators allowed. Note that an equivocator which is not allowed to state equivocate is essentially no different than the corresponding regular component.

\begin{definition}[Limited state equivocation model, \coqref{VLSM.Core.Equivocators.LimitedStateEquivocation}{equivocators_limited_equivocations_vlsm}]
The \textbf{$t$-limited state-equivocation model of $\{\mathcal{V}_i\}_{i=1}^n$} is the constrained VLSM composition in which we replace each component with its equivocating VLSM  under the composition constraint that components may only receive messages that have been sent in the current trace and the total weight of the equivocators does not exceed $t$. Formally, we have
\begin{align*}
\mathcal{V}_{{s\_\Eqv}}^{<t} = \Bigr({\sum_{i=1}^n} \mathcal{V}_i^e \Bigr) \Bigr|_{\varphi_{s\_\Eqv}^{<t}} &= (L,S, S_0, M, M_0,\tau, \valid \wedge {\varphi_{s\_\Eqv}^{<t}}), \\
\varphi_{s\_\Eqv}^{<t}( \iota , \langle \gamma_1,\mydots,\gamma_n\rangle,m) &= \sentM{\gamma_{\senderM{m}}}{m}\ \wedge \hspace{-.2cm} \sum_{\substack{k=1  \\ 1 < \len(\gamma_k')}}^n \hspace{-.2cm}\weight(k) < t,
\end{align*}
with $\langle \gamma_1',\mydots,\gamma_n'\rangle = \tau^s( \iota , \langle \gamma_1,\mydots,\gamma_n\rangle,m)$.
\end{definition}

Before we continue, let us point out some connections between the fixed-set and the weight-limited state equivocation models. First, note that equivocating traces have no message-equivocation and so they constitute a proof of validity of all their messages. Therefore, if the sum of weights of a set of indices $E$ is limited by the threshold $t$, it follows that any valid trace of $\mathcal{V}_{s\_\Eqv}^E$ is also a valid trace of $\mathcal{V}_{s\_\Eqv}^{<t}$, simply by replacing regular nodes with corresponding equivocating nodes which are not allowed to equivocate (\coqref{VLSM.Core.Equivocators.LimitedEquivocationSimulation}{equivocators_Fixed_incl_Limited}).
Conversely, given a valid trace $\tr$  of $\mathcal{V}_{s\_\Eqv}^{<t}$ whose last state is $\gamma = \langle \gamma_1,\mydots,\gamma_n \rangle$, by an argument similar to the above one (i.e., replacing equivocating nodes which are not allowed to equivocate with regular nodes), $\tr$ is a valid trace of $\mathcal{V}_{s\_\Eqv}^E$, where $E$ is the set of proper equivocators of $\tr$ which can be computed as $E = \{ i \ |\ 1 < \len(\gamma_i) \}$ (\coqref{VLSM.Core.Equivocators.LimitedStateEquivocation}{equivocators_limited_valid_trace_is_fixed}).

\subsubsection{The message equivocation model}

Based on the fixed-set message equivocation model, we can define the collection of traces with weight-limited message equivocation by taking the union of all valid traces for $\mathcal{V}^E_{m\_\Eqv}$ for all subsets $E$ whose weight is limited by $t$. We call them {\em traces under $t$-limited equivocation}. It is relatively easy to see that these traces correspond to subtraces of the weight-limited state equivocation model, by the following argument:
\begin{list}
{\arabic{cont}.}{\usecounter{cont}}
\item For a valid trace $\tr_m$ of $\mathcal{V}^E_{m\_\Eqv}$ with a subset $E$ whose weight is limited by $t$, by Theorem~\ref{fixed-set-equivocation}, there is a valid trace $\tr_s$ of $\mathcal{V}^E_{s\_\Eqv}$ corresponding to $\tr$. By the above remark, $\tr_s$ is also valid in $\mathcal{V}_{s\_\Eqv}^{<t}$ (\coqref{VLSM.Core.Equivocators.LimitedEquivocationSimulation}{limited_equivocators_finite_valid_trace_init_to_rev}). 

\item Conversely, given a valid trace $\tr_s$ of $\mathcal{V}_{s\_\Eqv}^{<t}$, let $\gamma = \langle \gamma_1,\mydots,\gamma_n \rangle$ be its final state, and let  $$E = \{ i\ |\ 1 < \len(\gamma_i) \}$$ be the equivocating indices in $\tr_s$. We have that the weight of $E$ is limited by the threshold, and, by one of the above remarks, $\tr_s$ is also valid in $\mathcal{V}_{s\_\Eqv}^E$. By Theorem~\ref{fixed-set-equivocation}, there is a valid trace $\tr_m$ of $\mathcal{V}^E_{m\_\Eqv}$ corresponding to $\tr_s$. Since $E$ is of limited weight, equivocation in $\tr_m$ is limited  (\coqref{VLSM.Core.Equivocators.LimitedStateEquivocation}{equivocators_limited_valid_trace_projects_to_fixed_limited_equivocation}). 
\end{list}
In the remainder of this section we will define a VLSM whose valid traces are precisely the traces with limited equivocation described above.

Expressing weight-limited equivocation as a constraint for the composition of regular components is problematic as such a constraint must detect the amount of equivocation encountered by only looking at the states of the individual components.
To understand why this is a non-trivial task, consider the following example.

\begin{example}
{\normalfont
Given an initial state $\langle \sigma_1, \sigma_2 \rangle$ and two transitions, one on component $1$, receiving nothing, transitioning to $\sigma'_1$ and producing $m_1$ and one on component $2$, receiving $m_1$, transitioning to $\sigma'_2$ and producing nothing.  After both transitions have occurred, the new state is $\langle \sigma_1', \sigma_2'\rangle$. If the transitions occurred in the order described above, there should be no equivocation. However, if they occurred in the reverse order, component $1$ should be considered as equivocating. Hence in some traces the weight of $\langle \sigma_1', \sigma_2'\rangle$ should be $0$, while in others it should be the weight of component $1$.
}
\end{example}

In what follows, we will present a simple model for weight-limited message equivocation based on annotating states with the set of equivocators observed in the current trace so far, which makes the task of detecting the equivocators in a state trivial.

\begin{definition}[Limited message equivocation model, \coqref{VLSM.Core.Equivocation.MsgDepLimitedEquivocation}{full_node_limited_equivocation_vlsm}]\label{t-weight-mesage-equivocation}
The \textbf{$t$-limited message equivocation model of $\{\mathcal{V}_i\}_{i=1}^n$} is obtained from the free composition $\sum_{i = 1}^n{V_i}$ by annotating the states (of the free composition) with sets of equivocators, updating those sets during transitions by adding the sender of a message if the message is an equivocation, and further constraining the validity constraint to only accept inputs that lead to states whose sets of equivocators are of limited weight. Formally, we have
$$\mathcal{V}_{m\_\Eqv}^{<t} = (L^{<t}, S^{<t}, S^{0,<t}, M^{<t}, M^{0,<t}, \tau^{<t}, \valid^{<t}),$$ where $L^{<t}$ is the same set of labels as for the free composition,
$S^{<t} = \{\langle \sigma,E\rangle\ |\ \sigma \in S \mbox{ and } E \subseteq \{1,\mydots,n\} \}$ consists of pairs of states of the free composition and sets of indices, $S^{<t}_0 = \{\langle \sigma,\emptyset\rangle\ |\ \sigma \in S_0\}$, $M^{<t}$ and $M^{<t}_0$ are the same set of (initial) messages as for the free composition, the transition function is defined as $\tau^{<t}(\iota,\langle \sigma,E \rangle,m) = (\langle \sigma',E' \rangle, m')$, where $\tau(\iota,\sigma,m) =  (\sigma',m')$ and 
\vspace{-.2cm}
\begin{align*}
E' = \left\{\begin{array}{rl}
     E, & \mbox{ if } \sentM{\sigma}{m}  \\
     E\cup \{\senderM{m}\}, & \mbox{otherwise}
     \end{array}\right. ,
\end{align*}
while the validity constraint predicate is $$\valid^{<t}(\iota, \langle \sigma,E\rangle,m) = \valid(\iota,\sigma,m) \wedge (\weight(E') < t),$$ where $\tau^{<t}(\iota,\langle \sigma,E \rangle,m) = (\langle \sigma',E' \rangle, m')$ and $\valid$ is the validity constraint in the free composition.
\end{definition}

We call any valid trace of $\mathcal{V}_{m\_\Eqv}^{<t}$ a \textit{trace exposed to under $t$-limited equivocation behavior}.

\subsubsection{Equivalence between the state and message equivocation models}

As in the case of the fixed set message equivocation model, the {\em \fullNode}  is essential in ensuring the adequacy of the model because whenever the received message is an equivocation, the only (possibly new) equivocator must be its sender, as all its dependencies were already received (and equivocations accounted for).

\begin{theorem} Under \fullNode\  for all components,
\begin{enumerate}
    \item The trace reduct of a valid trace of the state equivocation model is a valid trace for the message equivocation model (\coqref{VLSM.Core.Equivocators.LimitedStateEquivocation}{equivocators_limited_valid_trace_projects_to_annotated_limited_equivocation}). 
    \item Each valid trace for the message equivocation model can be ``lifted'' to a valid trace for the state equivocation model such that its trace reduct  is the original trace  (\coqref{VLSM.Core.Equivocators.LimitedEquivocationSimulation}{equivocators_limited_valid_trace_projects_to_annotated_limited_equivocation_rev}). 
\end{enumerate}
\end{theorem}
\section{Reduction of Byzantine to Equivocating Behaviour} \label{Byzantine}

Traditionally, consensus literature has defined a Byzantine participant in a consensus protocol to be one with arbitrary behaviour \cite{lamport}.  Sometimes Byzantine nodes have a measure of control over the network, with the ability to delay, duplicate, or drop messages.  
In the VLSM framework, messages can be received at any time, and they may be received multiple times or not at all. 

\subsection{Definition of Byzantine component}

We can model a Byzantine component as a VLSM that can send or receive any message at any time. However,  we want Byzantine components to not be able to forge messages on behalf of other components.  We capture this through the validity constraint ensuring that  any message sent by a Byzantine component is attributed  to its sender (\coqref{VLSM.Core.ByzantineTraces.FixedSetByzantineTraces}{emit_any_signed_message_vlsm_machine}).

\begin{definition}[Byzantine component, \coqref{VLSM.Core.ByzantineTraces}{emit_any_message_vlsm}]\label{Byzantine-VLSM}
A \textbf{Byzantine component (of address $i$)} is a VLSM of the form $\mathcal{B}_i = (L,S,S_{0}, M, M_{0}, \tau, \valid)$, where the labels are the messages, i.e., $L = M$,
there is only one state which is also an initial one, i.e., $S = S_0 = \{s\}$, the set of messages and initial messages coincide, i.e., $M_0 = M$, the transition function ignores received messages and produces the label message, i.e., $\tau(m,s,m') = (s,m)$, for any $m,m'\in M$, while the validity constraint predicate is defined as
$\valid(m,s,m') = (\sender(m) = i)$.
\end{definition}

We will argue that validators in a VLSM composition do not distinguish between Byzantine components and equivocating components as defined in Section \ref{modelsOfEquivocation}. Therefore, when analysing the security of a protocol, it suffices to consider only equivocating components. We will analyse two scenarios: one in which the number of Byzantine components is fixed, and one in which the number of Byzantine components is limited by their weights.

Let $\{\mathcal{V}_i\}_{i=1}^n$  be an indexed set of VLSMs. We assume that each $\mathcal{V}_i$ satisfies the  \textit{Channel Authentication assumption} (Assumption \ref{channel-authentication}), the \textit{Sent assumption} (Assumption \ref{sent}),  and the \textit{ full node assumption} (Assumption \ref{full-node}).  

\subsection{Fixed subset of Byzantine components}
Let us fix a subset $B \subseteq \{1,\mydots,n\}$. We assume that, for any $i\in B$, each component $\mathcal{V}_i$ can be replaced with a Byzantine component  of address $i$, $\mathcal{V}_i^B$, defined as in Definition \ref{Byzantine-VLSM}. Note that the only ingredient $\mathcal{V}_i$ and $\mathcal{V}_i^B$ have in common is the set of messages $M$.
Assuming that the components $\mathcal{V}_i$ with $i\not \in B$ are protocol-abiding, we will enforce them to only receive messages seen in the current trace. Let $\mathit{\mathit{NonByz}}$ denote the non-Byzantine components, i.e.,  $\mathit{\mathit{NonByz}} = \{1,\mydots,n\}\setminus B$.  

Formally (\coqref{VLSM.Core.ByzantineTraces.FixedSetByzantineTraces}{non_byzantine_not_equivocating_constraint}), we define the constrained composition 
$$\mathcal{V}_{\mathit{Byz}}^B =\ \Bigr({\sum_{i=1}^n} \mathcal{V}_i' \Bigr) \Bigr|_{\varphi_{\mathit{Byz}}} = (L,S, S_0, M, M_0,\tau, \valid \wedge {\varphi_{\mathit{Byz}}}) $$
where, for any $1\leq i \leq n$, $\mathcal{V}_i' = \mathcal{V}_i$ if $i\in \mathit{NonByz}$ and $\mathcal{V}_i' = \mathcal{V}_i^B$ if $i\in B$, and
$$\varphi_{\mathit{Byz}}(\langle i,l\rangle, \langle s_1,\mydots,s_n\rangle,m) =  i \in {\mathit{NonByz}} \to \sentM{s_{\senderM{m}}}{m}.$$

\vspace{.1cm}
\noindent We call the traces in $\mathit{Tr}(\mathcal{V}_{\mathit{Byz}}^B)|_{\mathit{NonByz}}$ {\em traces exposed to $B$-fixed Byzantine behavior}.

\begin{theorem}[\coqref{VLSM.Core.ByzantineTraces.FixedSetByzantineTraces}{validator_fixed_non_byzantine_eq_fixed_non_equivocating}]\label{fixed-byzantine-fault-tolerance}
If the components from $\mathit{\mathit{NonByz}}$ are validators for the message equivocation model for the fixed subset of equivocators $B$,  $\mathcal{V}_{m\_\Eqv}^B$, then the traces exposed to $B$-fixed Byzantine behavior coincide with the traces exposed to $B$-fixed equivocation behavior, i.e., $$\mathit{Tr}(\mathcal{V}_{\mathit{Byz}}^B)|_{\mathit{NonByz}} = \mathit{Tr}(\mathcal{V}_{m\_\Eqv}^B)|_{\mathit{NonByz}}.$$
\end{theorem}
%

\subsection{Weight-limited subsets of Byzantine components}
Let us fix a {\em threshold} $t$. Then, a component $\mathcal{V}_i$ is a {\em validator for the $t$-limited equivocation model},  $\mathcal{V}_{m\_\Eqv}^{<t}$,  if for any $l\in L_i$, $s_i\in S_i$, and $m\in M \cup \{\nomessage\}$, if $\valid_i(l,s_i,m)$ holds, then there exists a valid state $\langle \sigma, E\rangle$ of $\mathcal{V}_{m\_\Eqv}^{<t}$ such that the $i^{th}$ component of $\sigma$ is $s_i$, $m$ is valid in $\mathcal{V}_{m\_\Eqv}^{<t}$, and $\valid_{<t}(\langle i,l\rangle, \langle \sigma,E\rangle,m)$ holds.

\begin{theorem}[\coqref{VLSM.Core.ByzantineTraces.LimitedByzantineTraces}{msg_dep_validator_limited_non_equivocating_byzantine_traces_are_limited_non_equivocating}]\label{limited-byzantine-fault-tolerance}
If all components are validators for the $t$-limited message equivocation model  $\mathcal{V}_{m\_\Eqv}^{<t}$, then the possible behaviors of the non-equivocating components are the same under $t$-limited Byzantine behaviour as under $t$-limited equivocation behavior.
\end{theorem}

For both Theorem~\ref{fixed-byzantine-fault-tolerance} and Theorem~\ref{limited-byzantine-fault-tolerance}, we assumed that the non-Byzantine components satisfy the {\em \fullNode}.  We needed this assumption since we have used a very simple model for Byzantine components, assuming that they only satisfy the {\em  \channelAuthentication}. We believe that the \fullNode\  could be dropped if the Byzantine components would additionally satisfy the {\em \messageDependencies}  (and thus, the {\em unforgeability assumption}), and if the models of equivocation are updated accordingly.



%



\section{Concluding Remarks}

The goal of this work is to provide foundations for a theory of fault tolerance that can replace Byzantine fault tolerance analysis in settings where it is practical to have validators. We have shown that equivocation faults are exactly as expressive as Byzantine faults when it comes to their influence on validators. This result means that in an asynchronous network without guarantee of message arrival, Byzantine behavior is  precisely equivocation behavior as far as validators are concerned. Traces exposed to equivocation behavior thereby account for the effects of all possible hostile environments that validators might find themselves in.

In Theorems \ref{fixed-byzantine-fault-tolerance} and \ref{limited-byzantine-fault-tolerance} we assumed that components satisfy the {\em \fullNode}, as we used a very simple model for Byzantine components.
We believe that the \fullNode\  could be dropped if the Byzantine components would additionally satisfy the {\em \messageDependencies}  (and thus, the {\em unforgeability assumption}).

Limited equivocation does not guarantee that messages are delivered at all. Our full node assumption  insists that messages are received after their dependencies, but no other assumption on the timing or order of the arrival of messages was made during the course of this investigation. This leaves it to later work to define, account for, and limit synchronization faults.
This invites us to explore distributed systems design in different adversarial settings. For example, we can have distinct and independent limits on equivocation faults and synchronization faults. 

We also showed that all components of an ELMO protocol are examples of equivocation limited validators. In future work, we will show more examples of equivocation limited validators, including consensus protocols that are safe and non-trivial. Further specifying consensus protocols that exhibit provable liveness and high performance in systems with limited synchronization faults is also left for future work.

\bibliographystyle{eptcs}
\bibliography{vlsm}

\begin{thebibliography}{10}
\providecommand{\bibitemdeclare}[2]{}
\providecommand{\surnamestart}{}
\providecommand{\surnameend}{}
\providecommand{\urlprefix}{Available at }
\providecommand{\url}[1]{\texttt{#1}}
\providecommand{\href}[2]{\texttt{#2}}
\providecommand{\urlalt}[2]{\href{#1}{#2}}
\providecommand{\doi}[1]{doi:\urlalt{https://doi.org/#1}{#1}}
\providecommand{\eprint}[1]{arXiv:\urlalt{https://arxiv.org/abs/#1}{#1}}
\providecommand{\bibinfo}[2]{#2}

\bibitemdeclare{conference}{bar}
\bibitem{bar}
\bibinfo{author}{Amitanand~S. \surnamestart Aiyer\surnameend},
  \bibinfo{author}{Lorenzo \surnamestart Alvisi\surnameend},
  \bibinfo{author}{Allen \surnamestart Clement\surnameend},
  \bibinfo{author}{Mike \surnamestart Dahlin\surnameend},
  \bibinfo{author}{Jean-Philippe \surnamestart Martin\surnameend} \&
  \bibinfo{author}{Carl \surnamestart Porth\surnameend} (\bibinfo{year}{2005}):
  \emph{\bibinfo{title}{BAR fault tolerance for cooperative services}}.
\newblock In: {\slshape \bibinfo{booktitle}{ACM symposium on Operating systems
  principles}}, pp. \bibinfo{pages}{45--58}.

\bibitemdeclare{misc}{VLSM13}
\bibitem{VLSM13}
\bibinfo{author}{Mihai \surnamestart Calancea\surnameend},
  \bibinfo{author}{Denisa \surnamestart Diaconescu\surnameend},
  \bibinfo{author}{Wojciech \surnamestart Kołowski\surnameend},
  \bibinfo{author}{Elaine \surnamestart Li\surnameend},
  \bibinfo{author}{Brandon \surnamestart Moore\surnameend},
  \bibinfo{author}{Karl \surnamestart Palmskog\surnameend},
  \bibinfo{author}{Lucas \surnamestart Peña\surnameend},
  \bibinfo{author}{Grigore \surnamestart Roșu\surnameend},
  \bibinfo{author}{Traian~Florin \surnamestart Șerbănuță\surnameend},
  \bibinfo{author}{Ioan \surnamestart Teodorescu\surnameend},
  \bibinfo{author}{Dafina \surnamestart Trufaș\surnameend},
  \bibinfo{author}{Jan \surnamestart Tušil\surnameend} \&
  \bibinfo{author}{Vlad \surnamestart Zamfir\surnameend}
  (\bibinfo{year}{2023}): \emph{\bibinfo{title}{{VLSM version 1.3}}}.
\newblock \bibinfo{note}{\url{https://doi.org/10.5281/zenodo.10391528}}.

\bibitemdeclare{conference}{clement}
\bibitem{clement}
\bibinfo{author}{Allen \surnamestart Clement\surnameend},
  \bibinfo{author}{Flavio \surnamestart Junqueira\surnameend},
  \bibinfo{author}{Aniket \surnamestart Kate\surnameend} \&
  \bibinfo{author}{Rodrigo \surnamestart Rodrigues\surnameend}
  (\bibinfo{year}{2012}): \emph{\bibinfo{title}{On the (limited) power of
  non-equivocation}}.
\newblock In: {\slshape \bibinfo{booktitle}{Symposium on Principles of
  Distributed Computing}}, pp. \bibinfo{pages}{301--308}.

\bibitemdeclare{book}{ScienceProg}
\bibitem{ScienceProg}
\bibinfo{author}{David \surnamestart Gries\surnameend} (\bibinfo{year}{1981}):
  \emph{\bibinfo{title}{The Science of Programming}}.
\newblock \bibinfo{publisher}{Springer}, \doi{10.1007/978-1-4612-5983-1}.

\bibitemdeclare{conference}{jaffe}
\bibitem{jaffe}
\bibinfo{author}{Alexander \surnamestart Jaffe\surnameend},
  \bibinfo{author}{Thomas \surnamestart Moscibroda\surnameend} \&
  \bibinfo{author}{Siddhartha \surnamestart Sen\surnameend}
  (\bibinfo{year}{2012}): \emph{\bibinfo{title}{On the price of equivocation in
  {Byzantine} agreement}}.
\newblock In: {\slshape \bibinfo{booktitle}{Symposium on Principles of
  Distributed Computing}}, pp. \bibinfo{pages}{309--318}.

\bibitemdeclare{article}{lamport}
\bibitem{lamport}
\bibinfo{author}{Leslie \surnamestart Lamport\surnameend},
  \bibinfo{author}{Robert \surnamestart Shostak\surnameend} \&
  \bibinfo{author}{Marshall \surnamestart Pease\surnameend}
  (\bibinfo{year}{1982}): \emph{\bibinfo{title}{The {Byzantine} Generals
  Problem}}.
\newblock {\slshape \bibinfo{journal}{ACM Transactions on Programming Languages
  and Systems}} \bibinfo{volume}{4}(\bibinfo{number}{3}), pp.
  \bibinfo{pages}{382--401}.

\bibitemdeclare{conference}{madsen}
\bibitem{madsen}
\bibinfo{author}{Mads~F. \surnamestart Madsen\surnameend} \&
  \bibinfo{author}{S{\o}ren \surnamestart Debois\surnameend}
  (\bibinfo{year}{2020}): \emph{\bibinfo{title}{On the subject of
  non-equivocation: defining non-equivocation in synchronous agreement
  systems}}.
\newblock In: {\slshape \bibinfo{booktitle}{Symposium on Principles of
  Distributed Computing}}, pp. \bibinfo{pages}{159--168}.

\bibitemdeclare{unpublished}{bitcoin}
\bibitem{bitcoin}
\bibinfo{author}{Satoshi \surnamestart Nakamoto\surnameend}
  (\bibinfo{year}{2014}): \emph{\bibinfo{title}{Bitcoin: A Peer-to-Peer
  Electronic Cash System}}.
\newblock \bibinfo{note}{\url{https://bitcoin.org/bitcoin.pdf}}.

\bibitemdeclare{misc}{Coq}
\bibitem{Coq}
\bibinfo{author}{\surnamestart {The Coq Development Team}\surnameend}
  (\bibinfo{year}{2023}): \emph{\bibinfo{title}{{The Coq Proof Assistant}}}.
\newblock \bibinfo{note}{\url{https://doi.org/10.5281/zenodo.1003420}}.

\bibitemdeclare{unpublished}{vlad-2019}
\bibitem{vlad-2019}
\bibinfo{author}{Vlad \surnamestart Zamfir\surnameend}, \bibinfo{author}{Nate
  \surnamestart Rush\surnameend}, \bibinfo{author}{Aditya \surnamestart
  Asgaonkar\surnameend} \& \bibinfo{author}{Georgios \surnamestart
  Piliouras\surnameend} (\bibinfo{year}{2019}):
  \emph{\bibinfo{title}{Introducing the ``Minimal {CBC Casper}'' Family of
  Consensus Protocols}}.
\newblock \bibinfo{note}{\url{https://github.com/cbc-casper/cbc-casper-paper}}.

\end{thebibliography}
\end{document}